\documentclass[prd,aps,letterpaper,twocolumn,superscriptaddress,preprintnumbers,nofootinbib,floatfix]{revtex4}
\pdfoutput=1
\usepackage{xcolor}
\usepackage{graphicx}
\usepackage{amsmath}
\usepackage{amsfonts}
\usepackage{amssymb}
\usepackage{rotating}
\usepackage{subfigure}
\usepackage{paralist} 
\usepackage{verbatim}
\usepackage{float}
\usepackage{soul} 
\usepackage{appendix}
\usepackage{natbib}
\usepackage{epsfig}
\usepackage{hyperref}
\hypersetup{colorlinks,linkcolor=red,urlcolor=blue,citecolor=blue}
\usepackage[utf8]{inputenc}
\usepackage[english]{babel}
\usepackage{tabularx}
\newcolumntype{C}{>{\centering\arraybackslash}X}
\usepackage{mathpazo}
\usepackage{tgpagella}
\usepackage{float}
\usepackage{listings}

\normalsize

\usepackage{booktabs}
\usepackage{array, multirow}

\usepackage{lipsum}

\definecolor{backcolour}{rgb}{0.95,0.95,0.92}
\lstdefinestyle{mystyle}{
    basicstyle=\ttfamily\small\color{black},
    keywordstyle=\ttfamily\small\color{blue},
    stringstyle=\ttfamily\small\color{purple},
    breakatwhitespace=true,         
    breaklines=true,  
    backgroundcolor=\color{backcolour},
    numbers=left,
    numbersep=3pt,
    showstringspaces=false,
    numberstyle=\ttfamily\color{gray}\scriptsize,
    }

\definecolor{cadetblue}{rgb}{0.37, 0.62, 0.63}

\newcommand{\NS}[1]{\textcolor{magenta}{#1}}

\definecolor{numcolor}{RGB}{185,0,255}

\newcommand{\num}[1]{\textcolor{black}{#1}}

\newcommand{\stonybrook}{Physics and Astronomy Department, Stony Brook University, Stony Brook, NY  11794, USA}
\newcommand{\perimeter}{Perimeter Institute for Theoretical Physics, Waterloo, ON N2L2Y5, Canada}
\newcommand{\waterloo}{Waterloo Centre for Astrophysics, University of Waterloo, Waterloo, ON, N2L 3G1, Canada}
\newcommand{\waterloodept}{Department of Physics and Astronomy, University of Waterloo, 200 University Ave W, N2L 3G1, Waterloo, Canada}
\newcommand{\DunlapAstro}{David A. Dunlap Dept of Astronomy and Astrophysics, University of Toronto, 50 St George Street, Toronto ON, M5S 3H4, Canada}
\newcommand{\DunlapInstitute}{Dunlap Institute for Astronomy and Astrophysics, University of Toronto, 50 St. George St., Toronto, ON M5S 3H4, Canada}

\begin{document}

\title{Current and Future Constraints on the Primordial Power Spectrum}

\author{Zachary~Cheslog}
\affiliation{\stonybrook}

\author{Emily~Finson}
\affiliation{\stonybrook}

\author{Amanda~MacInnis}
\affiliation{\stonybrook}

\author{Neelima~Sehgal}
\affiliation{\stonybrook}

\author{Niayesh~Afshordi}
\affiliation{\waterloo}
\affiliation{\waterloodept}
\affiliation{\perimeter}

\author{Simran~K.~Nerval} 
\affiliation{\DunlapAstro} 
\affiliation{\DunlapInstitute}

\author{Ren\'ee~Hlo\v{z}ek}
\affiliation{\DunlapAstro} 
\affiliation{\DunlapInstitute}

\begin{abstract}
The primordial scalar power spectrum provides a powerful window onto early-universe physics, including a broad class of inflationary scenarios. These models often predict a characteristic scalar spectral index ($n_s$), nonzero running ($\alpha_s$), or more general deviations from a pure power-law form.
We find that freeing the number of light relic species ($N_{\rm eff}$) and the sum of the neutrino masses ($\sum m_\nu$) broadens the $n_s$ error contours enough that key early-Universe models are no longer excluded. In particular, while both Starobinsky inflation and the Bi-thermal Big Bang model are ruled out at about the 95\% confidence level in the six-parameter $\Lambda$CDM model, we find that in the nine-parameter $\Lambda{\rm CDM} + \alpha_s + N_{\rm eff} + \sum m_\nu$ model, CMB data from \textit{Planck}, ACT, and SPT (CMB-PAS) combined with DESI~DR2 give \num{$n_s = 0.9758 \pm 0.0064$} and \num{$\alpha_s = 0.0080 \pm 0.0064$}, consistent with both models. Future CMB facilities will sharply improve constraints on $n_s$, $\alpha_s$, and, more generally, the binned primordial power spectrum $\mathcal{P}(k)$. For example, CMB-PAS constrains $e^{-2\tau}\mathcal{P}(k = 0.2\,{\rm Mpc}^{-1})$ to \num{0.5\%}, while SO-like and CMB-HD-like surveys would tighten this to \num{0.1\%}; CMB-HD would also extend this measurement to smaller, previously unprobed scales, constraining $e^{-2\tau}\mathcal{P}(k = 30\,{\rm Mpc}^{-1})$ to within a factor of \num{ten}. We release our forecast code and update the public CMB-HD likelihood and Fisher codes to support \texttt{CLASS} in addition to \texttt{CAMB}.
\end{abstract}

\maketitle

\section{Introduction}
\label{sec:intro}

Measurements of the cosmic microwave background (CMB) can probe the physics of inflation or other early Universe models in a number of ways.  For example, quantum fluctuations just prior to inflation, phase transitions, or topological defects could generate primordial gravitational waves that would be imprinted in the B-mode pattern of the CMB~{\cite{CMB-S4:2016ple}. A challenge of this measurement is that many viable early Universe models predict primordial gravitational wave amplitudes that may be too small to detect in the foreseeable future~\cite{Kamionkowski:2015yta}. One can also look for signatures of primordial non-Gaussianity either in the CMB directly~\cite{Planck:2019kim} or via large-scale, scale-dependent bias using the kinetic Sunyaev-Zel'dovich effect~\cite{Munchmeyer:2018eey}.  In this work, we focus on probing early Universe models by measuring the primordial power spectrum of matter fluctuations. Many early Universe models have precise predictions for the amplitude and shape of the primordial power spectrum, and this primordial spectrum is imprinted in the primordial CMB as well as in the gravitational lensing of the CMB.  \\

The primordial power spectrum can be modeled by a power law with an amplitude ($A_s$) and slope ($n_s$), and additionally, with some change in the slope ($\alpha_\mathrm{s} = d n_\mathrm{s} / d\ln k$); $n_s$ and $\alpha_s$ are called the scalar spectral index and the running of the scalar spectral index, respectively.  One can also allow complete freedom in the shape of the primordial spectrum and use the CMB to constrain the value of the primordial spectrum in individual $k$-mode bins ($\mathcal{P}(k)$)~\cite{Bridle:2003sa, peiris2010shape,  Dvorkin:2011ui, hlozek2012atacama, Hazra:2013eva, Planck:2018jri, Raffaelli:2025kew, AtacamaCosmologyTelescope:2025nti}.  Since many early Universe models have specific predictions for $n_s$, $\alpha_s$, and the general shape of the primordial spectrum, tight constraints on these parameters can confirm these models or rule them out.

Here, we present current constraints on $A_s$, $n_s$, $\alpha_s$, and a general binned $\mathcal{P}(k)$ from a combination of the latest CMB data from the \text{Planck} satellite~\cite{Planck:2018vyg}, the Atacama Cosmology Telescope (ACT)~\cite{AtacamaCosmologyTelescope:2025blo}, and the South Pole Telescope (SPT)~\cite{SPT-3G:2025bzu}.  We add to this the latest Baryon Acoustic Oscillation (BAO) measurements from the Dark Energy Spectroscopic Instrument (DESI)~\cite{DESI:2025zgx}.  We show these constraints while varying the other cosmological parameters that are the most degenerate with them.  We also forecast constraints on these parameters using specifications similar to those of the upcoming Simons Observatory (SO)~\cite{SimonsObservatory:2025wwn} and the proposed CMB-HD experiment~\cite{CMB-HD:2022bsz}.  

This work is organized as follows.  In Section~\ref{sec:data}, we describe the datasets considered.  Sections~\ref{sec:method-current} and~\ref{sec:method-projected} describe our methods to obtain the current and projected constraints, respectively. We present a sample of early Universe theory models in Section~\ref{sec:theory} and our current and projected constraints in the context of these theory models in Section~\ref{sec:results}.  In Section~\ref{sec:discussion}, we discuss and conclude.

\section{Datasets Considered}
\label{sec:data}

\subsection{Current Datasets}
\label{sec:current-data}

In this work, we analyze a joint dataset consisting of the latest CMB temperature, polarization, and lensing measurements from \text{Planck}, ACT, and SPT. We refer to this combination as CMB-PAS ({\it{Planck}}-ACT-SPT). To this, we add the latest BAO measurements from DESI.  Specifically, these data include:

\begin{itemize}
    \item The 2018 \textit{Planck} PR3 likelihood, including the high-$\ell$ $TT, TE,$ and $EE$ likelihoods and the low-$\ell$ $TT$ likelihood~\cite{Planck:2018vyg}. For low-$\ell$ $EE$ we use either the standard \texttt{SimAll} likelihood or the \texttt{Sroll2} likelihood, as specified below.  The latter is a reanalysis of the {\it{Planck}} HFI data using the improved \texttt{Sroll2} mapmaking algorithm~\cite{Delouis:2019bub, Pagano:2019tci}

    \item The ACT DR6 CMB-only (foreground marginalized) likelihood~\cite{AtacamaCosmologyTelescope:2025blo}

     \item The SPT-3G D1 temperature and polarization \texttt{candl} likelihood~\cite{SPT-3G:2025bzu}
    
    \item The ACT-{\it{Planck}}-SPT (APS) lensing likelihood (extended version)~\cite{ACT:2025qjh}
    
    \item The DESI BAO likelihoods for DR1 and DR2~\cite{DESI:DR1, DESI:2025zgx}
\end{itemize}
Below, we describe each dataset in more detail.\\

The {\it{Planck}} 2018 PR3 likelihood provides nearly full-sky measurements of temperature and polarization anisotropies~\cite{Planck:2018nkj, Planck:2018vyg, Planck:2019nip}.
To minimize the overlap with ACT and SPT data, we truncate the {\it{Planck}} high-$\ell$ multipole ranges to $\ell$ $\leq$ 1000 for $TT$ and $\ell$ $\leq$ 600 for $TE$ and $EE$, as was also done in~\cite{AtacamaCosmologyTelescope:2025blo}.  We also include the {\it{Planck}} low-$\ell$ $TT$ likelihood, and either the {\it{Planck}} low-$\ell$ $EE$ \texttt{Sroll2}~\cite{Delouis:2019bub, Pagano:2019tci} or \texttt{SimAll}~\cite{Planck:2019nip} likelihood when combining {\it{Planck}} data with only ACT or with ACT and SPT data, respectively. The use of \texttt{Sroll2} when combining with only ACT data follows the convention of the ACT DR6 analysis~\cite{AtacamaCosmologyTelescope:2025blo}. The use of \texttt{SimAll}, and more specifically, an effective prior on the optical depth, $\tau$, of $\tau = 0.051 \pm 0.006$ derived from the \texttt{SimAll} likelihood~\cite{Planck:2020olo}, when combining with both ACT and SPT data follows the convention of the SPT-3G D1 analysis~\cite{SPT-3G:2025bzu}. \\

We use the ACT Data Release 6 (DR6) CMB-only likelihood, which provides measurements of the temperature and polarization power spectra ($TT, TE, EE$) over nearly half the sky at 98, 150, and 220~GHz~\cite{AtacamaCosmologyTelescope:2025blo}. The ACT DR6 angular power spectra extend over a multipole range of 600 to 8500, with noise levels lower than those of {\it{Planck}} on small scales. The CMB-only likelihood isolates CMB anisotropies over the multipole range of 600 to 6500 and was created by marginalizing over the foreground and systematic parameter uncertainties obtained from the multi-frequency likelihood; this CMB-only likelihood reproduces the results obtained from the full multi-frequency likelihood~\cite{AtacamaCosmologyTelescope:2025blo}.
This likelihood also retains two nuisance parameters:  $A_\mathrm{act}$, which is the overall temperature calibration, and $P_\mathrm{act}$, which is the overall polarization efficiency~\cite{AtacamaCosmologyTelescope:2025blo}. \\

We also include the SPT-3G Data Release 1 (D1) CMB-only likelihood \texttt{SPT-lite}~\cite{SPT-3G:2025bzu}. This likelihood provides $TT, TE$, and $EE$ spectra from observations of 4\% of the sky over the multipole range $\ell \in (400, 4000)$. In this CMB-only likelihood, foreground contributions have been marginalized over, and only a small number of calibration parameters ($A_\mathrm{cal}$ and $E_\mathrm{cal}$) are retained. SPT-3G D1 provides high signal-to-noise polarization measurements at small scales, which complement ACT measurements and improve constraints on damping tail physics and lensing smoothing.\\

For CMB lensing, we include the ACT-{\it{Planck}}-SPT (APS) lensing likelihood~\cite{ACT:2025qjh}, consisting of the power spectrum of the reconstructed lensing potential ($C^{\phi\phi}_{L}$), derived from a joint analysis of ACT DR6 lensing~\cite{ACT:2023dou, ACT:2023ubw, ACT:2023kun}, Planck PR4 lensing~\cite{Planck:2020olo,Carron:2022eyg}, and SPT MUSE analysis of their main-field-2-year polarization-only data~\cite{SPT-3G:2024atg}. This combined data set represents the most precise CMB lensing measurement to date, achieving a lensing detection with a signal-to-noise ratio of 61~\cite{ACT:2025qjh}. We use the extended variant of this likelihood, which includes lensing data from ACT DR6 containing multipoles $40 \leq L \leq 1300$, Planck PR4 lensing data from $8 \leq L \leq 400$, and SPT MUSE data over the range $20 \leq L \leq 3000$~\cite{Carron:2022eyg, ACT:2025qjh, SPT-3G:2024atg}. \\

We also include BAO measurements from DESI for both Data Release 1 and 2 (DR1 and DR2, respectively), which provide high precision constraints on the late-time expansion history using spectroscopic galaxy and quasar samples~\cite{DESI:DR1, DESI:2025zgx}. DESI DR1 represents the first cosmological results from DESI, delivering distance measurements across multiple tracer populations and redshift bins. DR2 significantly improves upon DR1 through increased survey volume, higher tracer densities, and improved control of observational systematics, which result in tighter constraints on cosmological distances. While DR2 supersedes DR1 in precision, we include both data sets to assess the impact of these improvements and to verify the robustness and consistency of our constraints across DESI releases.

\subsection{Upcoming and Future Datasets}
\label{sec:future-data}

In addition to obtaining constraints on the primordial power spectrum from current datasets, we forecast the constraints that can be obtained from future CMB and CMB lensing data. For this, we use mock data and covariance matrices that simulate CMB and CMB lensing data expected from SO-like~\cite{SimonsObservatory:2018koc,SimonsObservatory:2025wwn} and CMB-HD-like~\cite{Sehgal:2019ewc,CMB-HD:2022bsz} experiments, as well as the full DESI BAO dataset~\cite{DESI:2016fyo}.  We describe each mock dataset in more detail below.

The mock CMB data consist of CMB $TT, TE, EE$, and $BB$ spectra, plus the CMB lensing convergence spectrum $\kappa\kappa$. A sky area of 60\% is assumed for both SO and CMB-HD, and we model only the 90 and 150~GHz frequency channels, assuming the other channels will mainly be used to constrain foregrounds. For both SO and CMB-HD, we assume that polarization noise levels are $\sqrt{2}$ times larger than temperature noise levels and that temperature and polarization instrumental noise are uncorrelated.\\

For an SO-like experiment, we only model the SO LAT for the enhanced SO survey and assume temperature noise levels and beam sizes of 3.8 $\mu$K-arcmin noise and a 2.2 arcmin beam at 90~GHz, and 4.1 $\mu$K-arcmin noise and a 1.4 arcmin beam at 150~GHz~\cite{SimonsObservatory:2018koc}. We assume CMB and CMB lensing spectra have $\ell_\mathrm{min} = 30$ and $L_\mathrm{min} = 30$, respectively.  For $TT$ and $\kappa\kappa,~\ell_\mathrm{max} = L_\mathrm{max} = 3,000$; for $TE, EE$, and $BB,~\ell_\mathrm{max} = 5,000$.  

For a CMB-HD-like experiment, we assume temperature noise levels and beam sizes of 0.7 $\mu$K-arcmin noise and 0.42 arcmin beam at 90~GHz, and 0.8 $\mu$K-arcmin noise and 0.25 arcmin beam at 150~GHz~\cite{CMB-HD:2022bsz}.  The CMB-HD temperature noise also includes estimated residual extragalactic foregrounds from a simulation-based foreground-removal analysis~\cite{MacInnis2026}. For $TT, TE, EE$, and $\kappa\kappa$, we assume
$\ell_\mathrm{min} = L_\mathrm{min} = 30$, and  $\ell_\mathrm{max} = L_\mathrm{max} = 20,000$. 
(Technically, we assume CMB-HD will only measure $\ell > 1000$, and will use SO noise levels for $\ell < 1000$; however, since SO will be sample-variance limited for $30 < \ell < 1000$, we can just use CMB-HD $TT$, $TE$, and $EE$ noise in that range.)  For $BB$, we assume SO $BB$ noise levels for $30 < \ell < 1000$ and CMB-HD noise levels for $1000 < \ell < 20,000$, as was done in~\cite{macinnis2024}.  For the $\kappa\kappa$ spectra, we combine $EE$, $EB$, $TE$, and $TB$ lensing estimators, restricting the temperature data to multipoles below $\ell_\mathrm{max}^T = 5{,}000$ to avoid high-$\ell$ foreground contamination~\cite{vanEngelen:2013rla}; we further add the $TT$ estimator, where, for lensing multipoles $L < 5000$, we again impose $\ell_\mathrm{max}^T = 5{,}000$, while for $L > 5000$, we adopt a previous simulation-based estimate of the $TT$ lensing noise~\cite{Han:2021vtm}, scaled as discussed in~\cite{MacInnis2026}.\\

All the CMB forecasts in this work assume lensed CMB spectra, as opposed to delensed spectra, because we compare forecasts using \texttt{CAMB} and \texttt{CLASS} to check for consistency, and standard \texttt{CLASS} does not have the option to create delensed spectra.  The exception is Table~\ref{tab:baryonic}, where we use \texttt{CAMB} delensed spectra and a corresponding covariance matrix when exploring CMB-HD cosmology constraints while also varying baryonic physics parameters. \\

The mock BAO data consist of distance ratio measurements $r_s/d_{V}(z)$, where $r_s$ is the comoving sound horizon at the end of the baryon drag epoch, and $d_{V} (z)$ is given by:
    \begin{equation}
        d_V(z) \equiv \left[(1+z)^2 d_A^2(z)\frac{cz}{H(z)}\right]^{1/3},
        \label{BAOdv}
    \end{equation}
where $H(z)$ represents the expansion rate of the universe at redshift $z$, and $d_{A}(z)$ is the angular diameter distance to redshift $z$~\cite{DESI:2016fyo}. For DESI mock data, we use redshift ranges and projected uncertainties given by tables 2.3 (for $z \in [0.65, 1.85]$) and 2.5 (for $z \in [0.05, 0.45]$) in~\cite{DESI:2016fyo}, assuming a DESI experiment that covers 14,000 square degrees of the sky~\cite{DESI:2016fyo, macinnis2024}.\\

The covariance matrices for both the mock BAO and CMB data are the v1.2 covariance matrices available on the CMB-HD GitHub\footnote{\url{https://github.com/CMB-HD/hdMockData}}, and their calculations are described in~\cite{MacInnis2026}. To summarize, the CMB covariance matrix is the joint covariance of lensed $TT, TE, EE, BB$ plus lensing $\kappa\kappa$ spectra, including off-diagonal elements that account for lensing-induced correlations between the different spectra. The BAO covariance matrix uses only Gaussian/diagonal elements, whose variance is $\sigma_i^2$ for distance ratio measurement $r_s/d_{V}(z_i)$ in the $i^\mathrm{th}$ redshift bin centered at $z_i$.

\section{Method to Obtain Current Constraints} \label{sec:method-current}

In this section, we describe how we use the likelihoods from the current datasets described above to extract cosmological parameters. \\ 

The theoretical CMB and CMB lensing power spectra are computed using the Einstein-Boltzmann code \texttt{CAMB}~\cite{Lewis:1999bs}. We use the same \texttt{CAMB} accuracy settings
that were employed in the ACT DR6 parameter analysis~\cite{AtacamaCosmologyTelescope:2025nti} including the recombination model CosmoRec~\cite{chluba:cosmorec_2, chluba:cosmorec_1}.
The CMB theoretical predictions are obtained by varying the six $\Lambda$CDM parameters: the density of dark matter, $\Omega_{c}h^2$, the density of baryonic matter, $\Omega_{b}h^2$, the acoustic scale, 100$\theta_{MC}$, the optical depth at reionization, $\tau$, the amplitude of primordial scalar perturbations, $A_s$, and the power-law spectral index of primordial scalar perturbations, $n_s$.
When using a power-law model of the primordial power spectrum, we also vary $\alpha_\mathrm{s} = d n_\mathrm{s} / d\ln k$, which is the running of the scalar spectral index given by
\begin{equation}
    P_{\mathcal{R}}(k) = A_s \left( \frac{k}{k_0} \right)^{n_s-1+\frac{1}{2}\alpha_s \ln(k/k_0)}
    \label{eq:P(k)}.
\end{equation}
Here, $k_0$ is a pivot scale that we take to be $0.05~\mathrm{Mpc}^{-1}$~\cite{Lanusse:2014sra}.  

In addition, we vary the effective number of relativistic species, $N_{\mathrm{eff}}$, and the sum of the neutrino masses, $\sum m_\nu$, due to the degeneracies of these parameters with $n_s$ and $\alpha_s$.  When varying $\sum m_\nu$, we use a model with three massive degenerate neutrinos, as was done in~\cite{Planck:2018vyg,ACT:2023kun,AtacamaCosmologyTelescope:2025nti}.  This more closely matches the realistic neutrino cases of a normal or inverted mass hierarchy, as discussed in~\cite{Lesgourgues:2006nd} and~\cite{CORE:2016npo}; concentrating all of $\sum m_\nu$ in a single neutrino causes the transition of neutrinos from relativistic to non-relativistic to occur earlier, increasing the suppression of the small-scale matter power spectrum.

We also explore changing the model of dark energy motivated by the latest DESI results~\cite{DESI:2025zgx}.  We vary the parameters $w_0$ and $w_a$, which are the parameters of the Chevallier-Polarski-Linder (CPL) dark energy equation of state 
$w(z) = w_0 + w_a(1-a)$ where $a=(1+z)^{-1}$~\cite{Linder:2002et, Chevallier:2000qy}.  For the $w_0w_a$CDM evolving dark energy model, we use the Parameterized Post-Friedmann parameterization (referred to in \texttt{CAMB} as \texttt{DarkEnergyPPF}). This PPF description of dark energy allows $w$ to cross the phantom divide $(w<-1)$, where non-PPF models of dark energy may encounter numerical instability when $(w<-1)$~\cite{Fang_2008}. (The default \texttt{CAMB} code will not allow values of $w_0$ and $w_a$ that would result in $(w<-1)$, as it uses a non-PPF fluid dark energy model~\cite{Lewis:1999bs}.) \\

To obtain BAO theory predictions, we use \texttt{CAMB}, given the accuracy settings and cosmological parameters mentioned above, to compute the sound horizon at the drag epoch $r_s$, the comoving angular diameter distance $D_M(z)$, and the Hubble distance $D_H(z)$.
The BAO likelihood then constructs the observables $D_H/r_s$, $D_M/r_s$, and the volume-averaged distance ratio, $D_V/r_s$~\cite{Torrado:2020dgo, DESI:2025zgx}. \\

To obtain parameter constraints, we compare theory predictions to data by running Markov Chain Monte Carlo (MCMC)
chains with \texttt{Cobaya}~\cite{Torrado:2020dgo} using the built-in MCMC sampler employing the Metropolis-Hastings algorithm~\cite{Lewis:cobaya_mcmc_1,Lewis:cobaya_mcmc_2}.  We use the same priors on the cosmological parameters that were used in the ACT DR6 parameter analysis~\cite{AtacamaCosmologyTelescope:2025nti}; the exception is replacing the \texttt{Sroll2} low-$\ell$ $EE$ likelihood with a \texttt{SimAll}-derived effective Gaussian prior on $\tau$ of $\tau = 0.051 \pm 0.006$ when combining ACT and SPT data, as was done in~\cite{SPT-3G:2025bzu} and mentioned in Section~\ref{sec:data}. 
The priors on the foreground and systematic parameters for ACT, SPT, and {\it{Planck}} are left at their default values. A burn-in fraction of 50\% is removed from the MCMC chains; this disregards the first half of the chains to eliminate bias from arbitrary starting points and to ensure that the chain has stabilized.  The reduced Gelman-Rubin statistic ($R-1$)~\cite{gelman-rubin_1, gelman_rubin_2} is calculated after the burn in is removed to check for convergence; we consider an $R-1$ of less than 0.01 to be converged. \\

We also allow the primordial power spectrum of scalar perturbations to have a non-power-law shape and obtain constraints on a more general binned power spectrum, $\mathcal{P}(k)$, as done in many previous analyses~\cite{Bridle:2003sa, Guo:2011re, Aich:2011qv, hlozek2012atacama, Hunt:2013bha,  dePutter:2014hza, Miranda:2015cea, Hazra:2016fkm, Obied:2017tpd, Planck:2018jri, AtacamaCosmologyTelescope:2025nti,  Raffaelli:2025kew}.  To do this, we choose $k$ bins and prior ranges roughly following the ACT DR6 analysis~\cite{AtacamaCosmologyTelescope:2025nti}; however, we choose to use wider $k$ bins than in that work so that our higher $k$ bins are less correlated.  In particular, we define seven bins, with the first bin spanning 0.0000562 to 0.00367 Mpc$^{-1}$ and the remaining six bins covering the range 0.00367 to 0.571 Mpc$^{-1}$ with equally spaced bins in log space.  These bins cover the same range of wavenumbers as the 30 bins used in~\cite{AtacamaCosmologyTelescope:2025nti}.  We vary $e^{-2\tau}\mathcal{P}(k)$ in each bin as was done in~\cite{AtacamaCosmologyTelescope:2025nti}; the factor of $e^{-2\tau}$ is added because of the degeneracy between $\tau$ and the primordial power spectrum.
Priors on $e^{-2\tau}\mathcal{P}(k)$ in each $k$ bin are adapted from the priors given in Table 5 of Appendix C of~\cite{AtacamaCosmologyTelescope:2025nti}. For our first bin, we select the prior of the first ACT DR6 bin in order to be conservative.  For the following bins, we choose as a prior the prior from Table 5 of~\cite{AtacamaCosmologyTelescope:2025nti} of the $k$ bin best aligned with our bin center. This yields conservative priors since our bins are wider than those of that work, and thus, they should have smaller errors.  In Table~\ref{tab:P(k)}, we list the $k$ bin centers and prior ranges used in this work.

\begin{table}[t]
    \begin{center}
      \begin{tabular}{c@{\hskip 0.5em} | c@{\hskip 0.5em}}
      \toprule
      \hline
      $k$  [Mpc$^{-1}$] & ~~$e^{-2\tau}\mathcal{P}(k)$ Prior Range $\times 10^9$  \\ \hline
      0.0018 & $ 0.00 - 15.0$ \\ \hline
      0.0060 & $ 0.50 - 8.00$ \\ \hline
      0.0141 & $ 1.40 - 2.40$ \\ \hline
      0.0327 & $ 1.68 - 2.14$ \\ \hline
      0.0759 & $ 1.70 - 2.00$ \\ \hline
      0.176 & $ 1.04 - 2.55$ \\ \hline
      0.408 & $ 0.00 - 8.46$ \\ \hline
      0.947 & --- \\ \hline
      2.20  & --- \\ \hline 
      5.09 & --- \\ \hline 
      11.8 & --- \\ \hline 
      27.4 & --- \\ \hline 
    \end{tabular}
    \caption{We allow the primordial power spectrum of scalar perturbations to have a non-power-law shape and obtain constraints on a more general binned power spectrum, $\mathcal{P}(k)$. Here we give the central-bin wavenumber and prior ranges used for sampling $e^{-2\tau}\mathcal{P}(k)$ for each $k$ bin when obtaining constraints from P-ACT-LB and CMB-PAS data.  We also give the central-bin wavenumbers used when obtaining Fisher forecasts for SO and CMB-HD, as described in Section~\ref{sec:method-projected}. From the second bin onward we use equally spaced bins in log space.}
    \label{tab:P(k)}
    \end{center}
\end{table}

We use the \texttt{BinnedPk} code from \texttt{Cobaya} to extract $\mathcal{P}(k)$ for each bin and pass these to \texttt{CAMB}.\footnote{Note that this code defaults to equal log space bins unless modified.} We use the cubic spline
interpolation method within \texttt{CAMB} via \texttt{initialpower.SplinedInitialPower} to construct
the initial power spectrum from our binned values. \texttt{CAMB} then calculates the relevant theory (CMB spectra plus BAO) to be used in the likelihood.  

In addition to these binned $\mathcal{P}(k)$ values, we vary the four additional $\Lambda$CDM parameters: $\Omega_b h^2$, $\Omega_ch^2$, $\tau$, and $\theta_\mathrm{MC}$. (We no longer vary $n_\mathrm{s}$ and $A_\mathrm{s}$ since the binned $\mathcal{P}(k)$ values replace them.)  We found that replacing the \texttt{Sroll2} low-$\ell$ $EE$ likelihood with an effective Gaussian prior on $\tau$ of $\tau = 0.0566 \pm 0.0058$, as sometimes done in~\cite{AtacamaCosmologyTelescope:2025nti}, sped up the convergence of the binned $\mathcal{P}(k)$ when combining {\it{Planck}} and ACT data.  As mentioned above, we use an effective Gaussian prior on $\tau$ of $\tau = 0.051 \pm 0.006$ when combining {\it{Planck}}, ACT, and SPT data.  We confirmed using the original 30 $\mathcal{P}(k)$ bins from~\cite{AtacamaCosmologyTelescope:2025nti} and {\it{Planck}} plus ACT data that the use of either the \texttt{Sroll2} low-$\ell$ $EE$ likelihood or the effective Gaussian prior makes minimal difference to the binned $\mathcal{P}(k)$ errors and marginalized means; in particular, the former agrees to within 13\% and the latter to within 3\%.\\

To determine the convergence of the subsequent MCMC chains, we calculate the reduced Gelman-Rubin statistic for all varied parameters given by
\begin{equation}
R - 1 = \sqrt{\mathrm{max}(W^{-1} V)} - 1
\end{equation}
\noindent where $\mathrm{max}$ refers to the maximum eigenvalue of $W^{-1} V$, $V = \frac{n-1}{n} W + \frac{1}{n} B$, $n$ is the length of each chain, $B/n$ is the mean of the variance between chains, and $W$ is the mean within-chain variance. 
When performing binned power spectrum reconstructions, the bin amplitude parameters are highly correlated. The standard convergence computation within \texttt{getdist} struggles to compute the inverse of the within-chain covariance matrix, $W$, due to its highly correlated nature. 
To address this, we implement a modified routine to compute the Gelman-Rubin statistic, which uses the pseudo-inverse routine \texttt{np.linalg.pinv} that utilizes singular value decomposition.

\section{Method to Obtain Projected Constraints}
\label{sec:method-projected}

Upcoming and future CMB data, such as from the Simons Observatory~\cite{SimonsObservatory:2025wwn} and CMB-HD~\cite{CMB-HD:2022bsz}, will measure the CMB with higher sensitivity than current datasets.  CMB-HD will also measure the CMB with higher resolution than precursor surveys. We forecast constraints on cosmological parameters from an SO-like and CMB-HD-like experiment to assess the impact of these improved measurements. Here, we describe our forecast methods using the mock data and covariance matrices described in Section~\ref{sec:data}. \\

We employ a Fisher information matrix method to make the forecasts presented in this work and cross-check some of these results by running MCMC chains, as described in Appendix~\ref{sec:stabillity}. We follow the method described in~\cite{macinnis2024} for the Fisher matrix calculations, which we summarize below.  \\

{\it{CMB Fisher matrix:}} The CMB likelihood developed in~\cite{macinnis2024} is given by 
\begin{equation}
    -2 \ln \mathcal{L}_\mathrm{CMB}\left(\hat{C}_{\ell_b} | \vec{\theta} \right) = \sum_{\ell_b \ell_{b'}} \Delta C_{\ell_b}(\vec{\theta}) \mathbb{C}^{-1}_{\ell_b \ell_{b'}}  \Delta C_{\ell_{b'}}(\vec{\theta}),
\label{eq:likelihood}  
\end{equation}
where $\vec{\theta}$ are all the free parameters in the model, $\Delta C_{\ell_b}(\vec{\theta}) = \hat{C}_{\ell_b} - C_{\ell_b}(\vec{\theta})$, $\hat{C}_{\ell_b}$ is the binned data spectra at bin center $\ell_b$,  $C_{\ell_b}(\vec{\theta})$ is the binned theory spectra, and $\mathbb{C}_{\ell_b \ell_{b'}}$ is the binned covariance matrix of the data. 
We assume the likelihood is maximized for the fiducial parameters $\vec{\theta}_0$ and Taylor expand about $\vec{\theta}_0$. Then, the elements of the Fisher matrix are given by 
\begin{equation}
     F_{\alpha\beta}^\mathrm{CMB} = \sum_{\ell_b, \ell_b'} \left. \left(\frac{\partial C_{\ell_b}}{\partial \theta_\alpha} \mathbb{C}^{-1}_{\ell_b \ell_b'} \frac{\partial C_{\ell_b'}}{\partial \theta_\beta} \right) \right|_{\vec{\theta}_0}
\label{eq:fisher_elements}
\end{equation}
where $\alpha$ and $\beta$ are indices corresponding to the parameters.  We obtain the CMB $TT, TE, EE, BB,$ and the lensing $\kappa \kappa$ power spectra using either \texttt{CLASS} or \texttt{CAMB} and bin them with a binning matrix.  The derivatives of the spectra with respect to each parameter are calculated using a finite difference method: each parameter is varied both up and down by the step size listed in Table~\ref{tab:fisher}, holding the other parameters fixed to the fiducial values in Table~\ref{tab:fisher}, and the derivative is given by the difference of the varied spectra divided by two times the step size.\footnote{We note that Fisher errors have a small dependence on the fiducial parameter values. While we calculate Fisher derivatives using the fiducial model in Table~\ref{tab:fisher}, for ease of comparing current and projected constraints, we center projected constraints on CMB-PAS plus DESI DR2 marginalized mean values in Figures~\ref{fig:alpha_s-vs-n_s-forecasts} and~\ref{fig:hd_so_pas_triangle}. To quantify the impact of fiducial models, we recompute the Fisher matrix for the $\Lambda\mathrm{CDM}+\alpha_s+N_\mathrm{eff}+\sum m_\nu$ model with $\vec{\theta}_0$ set to the CMB-PAS plus DESI DR2 marginalized means of Table~\ref{tab:current-constraints} (using 0.056 for $\sum m_\nu$), and find that the forecast $1\sigma$ errors change by at most \num{2\%}, with the greatest change in $\Omega_b h^2$ and the rest of the parameter errors changing by less than \num{1\%}.} \\

{\it{BAO Fisher matrix:}} The BAO likelihood developed in~\cite{macinnis2024} is given by 
\begin{equation}
  -2\ln\mathcal{L_\mathrm{BAO}}(\hat{f_i}|\vec{\theta}) = \sum_i\frac{[\hat{f_i}-f_i(\vec{\theta})]^2}{\sigma^2_i}
  \label{eq:bao_like}
\end{equation}
where $f_i =\frac{r_s}{d_{V}(z_i)}$, $\hat{f_i}$ is the BAO data, which at redshift $z_i$ has a variance $\sigma^2_i$, and $f_i(\vec{\theta})$ is the BAO theory evaluated at parameters $\vec{\theta}$.  Using this BAO likelihood, the elements of the BAO Fisher matrix are given by
\begin{equation} 
    F_{\alpha\beta}^\mathrm{BAO} = \sum_i \left. \left(\frac{\partial f_i}{\partial \theta_\alpha} \frac{1}{\sigma^2_i} \frac{\partial f_i}{\partial \theta_\beta} \right) \right|_{\vec{\theta}_0},
\label{eq:fisherBAO}    
\end{equation}
where we sum over the redshifts $z_i$. We use either \texttt{CLASS} or \texttt{CAMB} to calculate the values of $\frac{r_s}{d_{V}(z_i)}$ for each redshift given by tables 2.3 and 2.5 in~\cite{DESI:2016fyo}. \\

The total Fisher matrix for both CMB and BAO combined is then calculated by summing the two, i.e.~$F = F^{\mathrm{BAO}} + F^{\mathrm{CMB}}$. The  covariance matrix of the parameters is then given by the inverse of the Fisher matrix. The diagonal elements of this parameter covariance matrix give the variances, $\sigma^2$, of each parameter.  \\

\begin{table}[t]
    \begin{center}
    \begin{tabular}{l@{\hskip 1.5em} c@{\hskip 1.5em} c c}
      \toprule
      \toprule
      Parameter & Fiducial  & Step Size & Prior
      \\
      \midrule
      $\Omega_\mathrm{b} h^2$\dotfill & $0.02237$ & 1\% & $[0.005,\ 0.1]$
      \\
      $\Omega_\mathrm{c} h^2$\dotfill & $0.1200$ & 1\% & $[0.001,\ 0.99]$
      \\
      $H_0$ [km s$^{-1}$ Mpc$^{-1}$]\dotfill & $67.36$ & 1\% & $[20,\ 100]$
      \\
      $\tau$\dotfill & $0.0544$ & 5\% & $0.054 \pm 0.005$
      \\
      $\ln(10^{10} A_\mathrm{s})$\dotfill & $3.044$ & 0.3\%\footnote{The step size of 0.3\%  corresponds to a step size of about 1\% on $A_\mathrm{s}$.} & $[2,\ 4]$
      \\
      $n_\mathrm{s}$\dotfill & $0.9649$ & 1\% & $[0.8,\ 1.2]$
      \\
      $\alpha_s$\dotfill & 0.00 & 0.01 & [-0.2, 0.2] 
      \\
      $w_0$\dotfill & -1.0 & 0.01\% & ---
      \\
      $w_a$\dotfill & 0.0 & 0.01 & ---
      \\
      $N_\mathrm{eff}$\dotfill & $3.044$ & 5\%\footnote{Note \texttt{CLASS} restricts how far $N_\mathrm{eff}$ can deviate below the standard value which we bypass (see Appendix~\ref{sec:accuracy}).} & $[0.05,\ 10]$
      \\
      $\sum m_\nu$ [eV]\dotfill & $0.06$ & 10\% & $[0,\ 5]$ \\
      $\log_{10}(\mathrm{T_{AGN}}/\mathrm{K})$\dotfill  & 7.8 & 0.05 & $7.8 \pm 0.005$ \\
      $A_\mathrm{kSZ}$\dotfill &1.0 & 0.1 & ---\\
      $n_\mathrm{kSZ}$\dotfill &0.0 & 0.01 & ---
      \vspace{1mm}
      \\
      \hline
      $100 \theta_\mathrm{MC}$\dotfill & $1.04071$ & 1\% & $[0.5,\ 10]$
      \\
      $100 \theta_*$\dotfill & $1.041697$ & 1\% & $[0.5,\ 10]$
      \\
      \bottomrule
    \end{tabular}
    \caption{The fiducial cosmological parameters, step sizes, and priors used for the forecasts in this work. The second column lists the fiducial value of each parameter; we use the Planck 2018~\cite{Planck:2018vyg} cosmological parameter values of the six $\Lambda$CDM parameters (first six rows), and the standard, theoretical values of the other parameters. 
    The values for $100\theta_\mathrm{MC}$ (used by \texttt{CAMB}) and $100\theta_*$ (used by \texttt{CLASS}) are obtained by passing the other parameters, including $H_0$, into \texttt{CAMB}/\texttt{\texttt{CLASS}} with our accuracy settings; our higher accuracy settings cause both to differ from the values obtained with \textit{Planck}~\cite{Planck:2018vyg} accuracy settings.
    The third column lists the step sizes used when varying each parameter in the Fisher matrix calculations, and the fourth column lists the priors used in the MCMC runs on the mock CMB and BAO data.  All the priors are uniform except for $\tau$ and $\log_{10}(\mathrm{T_{AGN}}/\mathrm{K})$ which are Gaussians; the $\tau$ and $\log_{10}(\mathrm{T_{AGN}}/\mathrm{K})$ priors are also applied to the Fisher estimates. The error on $\tau$ for the prior is taken from the error bar on $\tau$ from the CMB-PAS plus DESI DR2 data in the $\Lambda \mathrm{CDM} + \alpha_s$ model (see Table~\ref{tab:current-constraints}). The error bar on $\log_{10}(\mathrm{T_{AGN}}/\mathrm{K})$ is equivalent to 0.06\%, which is the same as used in~\cite{MacInnis:darkmatter, macinnis2024} and is anticipated from combined CMB-HD tSZ, kSZ, and lensing measurements. }
    \label{tab:fisher}
    \end{center}
\end{table}

In Table~\ref{tab:fisher}, we show the fiducial cosmological parameter values used for our forecasts (second column), as well as the step sizes used for the Fisher calculation (third column) and the priors used for the MCMC runs (fourth column).  The fiducial values for the six $\Lambda\mathrm{CDM}$ parameters are taken from the Planck 2018 results~\cite{Planck:2018vyg}, and we use standard $\Lambda\mathrm{CDM}$ theoretical values for the remaining parameters, plus a minimum sum of neutrino masses.  Since \texttt{CAMB} and \texttt{CLASS} use different definitions of the acoustic scale $\theta$ (\texttt{CLASS} uses $\theta_\star$ while \texttt{CAMB} uses $\theta_\mathrm{MC}$), we choose to vary $H_0$ directly for the forecasts presented in this work instead and give $100\theta_\mathrm{MC}$ and $100\theta_\star$ as derived parameters.  We obtain $\theta_\star$ and $\theta_\mathrm{MC}$ using \texttt{CLASS} and \texttt{CAMB}, respectively; due to our higher accuracy settings (described below and given in Listings~\ref{list:camb} and~\ref{list:class}), these values differ from those obtained with the {\it{Planck}} accuracy settings~\cite{Planck:2018vyg}.

Additionally, \texttt{CLASS} handles the $N_\mathrm{eff}$ parameter and the $w_0$ and $w_a$ parameters for the $w_0w_a$CDM model differently from \texttt{CAMB}. 

For $N_\mathrm{eff}$, \texttt{CLASS} supports two different inputs: \texttt{Neff} or \texttt{N\_ur}, where \texttt{N\_ur}~=~$N_\mathrm{eff} - 1.0131966(n_\mathrm{nu})$ and $n_\mathrm{nu}$ is the number of massive neutrinos (set to three for forecasts in this work)~\cite{lesgourgues2011, Diego_Blas_2011}. We choose to pass \texttt{N\_ur}, computing it from our $N_\mathrm{eff}$ value, since we observe better agreement between \texttt{CAMB} and \texttt{CLASS} spectra below $\ell=15,000$ when inputting \texttt{N\_ur} directly;  calculating \texttt{N\_ur} externally and then inputting it into \texttt{CLASS} yields agreement between \texttt{CAMB} and \texttt{CLASS} spectra to within 0.2\%, whereas inputting \texttt{Neff} into \texttt{CLASS} yields agreement to within 0.5\%.

In order to calculate spectra with the dark energy parameters $w_0$ and $w_a$ using \texttt{CLASS}, we note that by default, \texttt{CLASS} leaves \texttt{Omega\_Lambda} unspecified and sets \texttt{Omega\_fld}, the density of the fluid dark energy, and \texttt{Omega\_scf}, the density of a scalar field component, to zero for a universe with dark energy made up exclusively of a cosmological constant. To activate the fluid dark energy model instead, we set $\texttt{Omega\_Lambda}=0$ and leave \texttt{Omega\_fld} unspecified; \texttt{CLASS} then infers \texttt{Omega\_fld} from the closure equation ($\sum_i \Omega_i =1+\Omega_k$, where $\Omega_k$ is the curvature density, set to zero for this work).~\cite{lesgourgues2011, Diego_Blas_2011}.   When we do this, we set \texttt{use\_ppf = yes} and \texttt{fluid\_equation\_of\_state = CLP}, where ppf refers to the Parameterized Post-Friedmann parameterization~\cite{Fang_2008} and CPL refers to the Chevallier-Polarski-Linder (CPL) dark energy equation of state ($w(z) = w_0 + w_a(1-a)$)~\cite{Linder:2002et, Chevallier:2000qy}.  With these settings, we find poor agreement between \texttt{CAMB} and \texttt{CLASS} parameter error bars, disagreeing by as much as about \num{45}\%. Given the significant disagreement, we choose to show only \texttt{CAMB} forecasts for the $w_0w_a$CDM model. We select \texttt{CAMB} for our forecasts with this model to maintain consistency with what was used in~\cite{DESI:2025zgx, AtacamaCosmologyTelescope:2025nti}.

The step sizes are the same as those used in~\cite{macinnis2024}, with the exception of $\alpha_s$, $w_0$, and $w_a$, which were not considered in that work; for these, we determine the step sizes by testing the stability of the Fisher matrix as we vary them. We choose step sizes such that a slight increase or decrease in the step size does not produce a significant change in the Fisher matrix.\\

All the priors listed are uniform, except for the priors on $\tau$ and $\log_{10}(\mathrm{T_{AGN}}/\mathrm{K})$, which are Gaussian. This $\tau$ prior is also applied to our Fisher forecasts by adding the inverse of its variance, $\frac{1}{\sigma^2}$, to the $F_{\tau\tau}$ element of the Fisher matrix. The prior of $\sigma(\tau) = 0.005$ was chosen as it represents the uncertainty on $\tau$ from the CMB-PAS + DESI DR2 constraints on the $\Lambda \mathrm{CDM} + \alpha_s$ model (see Table~\ref{tab:current-constraints} below).  The baryonic physics prior of 0.06\% is also added to our Fisher forecasts when varying $\log_{10}(\mathrm{T_{AGN}}/\mathrm{K})$ in the model.  We do not list priors on $w_0$ and $w_a$ since we do not run MCMC forecasts while varying them. \\

{\it{CMB Accuracy Settings:}} These projected constraints rely on accurate theoretical predictions of the CMB and CMB lensing power spectra. The \texttt{CAMB} or \texttt{CLASS} accuracy settings used in current cosmological analyses, e.g., \cite{AtacamaCosmologyTelescope:2025nti}, are insufficient at the low noise levels and small scales  required for these forecasts.  Calculating theory power spectra with insufficient accuracy could lead to biases in the inferred cosmological parameter values.  Previous work has found \texttt{CAMB} accuracy settings that are sufficient for CMB-HD forecasts~\cite{macinnis2024, MacInnis:darkmatter}; however, this had not been done for \texttt{CLASS}.  In this work, we find a set of \texttt{CLASS} accuracy parameters that results in parameter biases of less than $0.5\sigma$ for CMB-HD, which we describe in detail in Appendix~\ref{sec:accuracy}.  The \texttt{CAMB} and \texttt{CLASS} accuracy settings used here are given in Appendix~\ref{sec:accuracy} in Listings~\ref{list:camb} and~\ref{list:class}, respectively.\\

Unless stated otherwise, all the Fisher forecasts presented in this work are calculated with \texttt{CLASS}.  As mentioned in Section~\ref{sec:data}, we also use lensed CMB spectra throughout, as opposed to delensed CMB spectra, since standard \texttt{CLASS} currently does not support delensing. Thus, our forecasts are mildly conservative.\footnote{We find that \texttt{CAMB} Fisher errors on parameters using CMB-HD-like delensed spectra are, at most, \num{6\%} smaller than when using lensed spectra for the 9-parameter $\Lambda\mathrm{CDM}+\alpha_s+N_\mathrm{eff}+\sum m_\nu$ model.} We also update the \texttt{hdfisher}\footnote{\href{https://github.com/CMB-HD/hdfisher}{https://github.com/CMB-HD/hdfisher}} and \texttt{hdlike}\footnote{\href{https://github.com/CMB-HD/hdlike}{https://github.com/CMB-HD/hdlike}} \texttt{GitHub} repositories to allow the calculation of Fisher matrices and the ability to run MCMC chains for CMB-HD using \texttt{CLASS}. \\

We verify that our parameter forecasts are stable and consistent across both Boltzmann codes (\texttt{CAMB} and \texttt{CLASS}) and both forecasting methods (Fisher estimates and MCMC), as shown in Appendix~\ref{sec:stabillity}.

\subsection{Method to Obtain Binned $\mathcal{P}(k)$ Forecasts}
\label{sec:method-projected-Pk}

We discuss our method for obtaining the binned primordial power spectrum, $\mathcal{P}(k)$, forecasts separately since we need to bypass the standard non-linear matter power spectrum code to achieve this.  For these forecasts, we use only \texttt{CAMB}.\\

We define 11 $k$ bins, covering the range 0.00367 to 38.3 Mpc$^{-1}$ with equally spaced bins in log space.  We use all 11 bins to forecast for CMB-HD, but only the first seven bins to forecast for an SO-like experiment; we find that including bins with $k>1$~Mpc$^{-1}$ for SO worsens the SO constraints for $k>0.07$~Mpc$^{-1}$ due to covariances between the bins at high $k$.  The first six of these bins for both CMB-HD and SO are the same as those we used for the P-ACT-LB and CMB-PAS analyses described in Section~\ref{sec:method-current}.  We list the bin centers in Table~\ref{tab:P(k)}, and we calculate the fiducial values of $\mathcal{P}(k)$ in each bin using equation~\ref{eq:P(k)}, with parameters set to their fiducial values in Table~\ref{tab:fisher}. \\

To calculate derivatives for the Fisher matrix, we use a step size of 5\% on each of the $\mathcal{P}(k)$ bins. We select this step size by testing several different step sizes and checking for stability in the resulting forecasts, as was done to select the step sizes for $\alpha_s$, $w_0$, and $w_a$.  A varied bin is represented by multiplying the fiducial power-law curve by one plus or minus the step size (i.e.~$1~\pm$ step) over the $k$ range of that bin. We densely sample this curve and pass it to \texttt{CAMB} as a function via \texttt{set\_initial\_power\_function}, which builds the cubic spline used internally. \\

When using a binned primordial power spectrum, \texttt{CAMB} requires the parameter \texttt{effective\_ns\_for\_nonlinear} to be passed to  the non-linear model \texttt{HMcode}; it also sets this parameter to a default value. This causes the lensing power spectrum not to reflect the changes caused by shifting the primordial bins up or down, especially for high $k$ modes where non-linear effects are important. Therefore, we instead employ the method used by~\cite{MacInnis:darkmatter} to compute the lensed CMB spectra ($TT, TE, EE, BB$) and the lensing convergence spectra ($\kappa \kappa$), as we describe below.  

Since the issue mentioned above only affects the lensed spectra, we obtain the unlensed spectra via \texttt{CAMB}, using the accuracy settings given in Listing~\ref{list:camb} and the binned primordial power spectra constructed using \texttt{SplinedInitialPower}. Following~\cite{Lewis:2006fu}, we obtain the lensing potential power spectrum by 
\begin{equation}
            C_{\ell}^{\phi \phi} = 4\int_{0}^{\chi_s} d\chi \left( \frac{\chi_s - \chi}{\chi^2\chi_s}\right)^2 P_\Psi\left( k = \frac{\ell + 1/2}{\chi}, z(\chi)\right).
            \label{eq:lensing_int}
\end{equation}
Here, the comoving wavenumber is in $\mathrm{Mpc}^{-1}$ and $k\approx (\ell+1/2)/(\chi(z))$. $\chi(z)$ is the comoving distance to redshift $z$ in Mpc~\cite{Lewis:2006fu, dodelson2021modern}, and $\chi_s = \chi(z_s)$ is the comoving distance to the last scattering surface at $z\approx 1100$. We obtain the lensing convergence power spectrum $C_{\ell}^{\kappa \kappa}$ by
\begin{equation}
            C_{\ell}^{\kappa \kappa} = \frac{[\ell(\ell+1)]^2}{4} C_{\ell}^{\phi \phi}.
            \label{eq:clkk}
\end{equation}
$P_\Psi(k,z)$ is the power spectrum of the three dimensional gravitational potential $\Psi(\mathbf{k},z)$, which is related to the non-linear matter power spectrum $P_m(k,z)$ by~\cite{Lewis:2006fu, Nguyen:2017zqu, dodelson2021modern}
\begin{equation}
            P_\Psi(k,z) = \left( \frac{3 \Omega_m H_0^2}{2c^2}\right)^2 \frac{(1+z)^2}{k^4}P_m(k,z).
            \label{eq:gravpower}
\end{equation}
Here $H_0$ is the Hubble parameter today, and $\Omega_m$ is the matter density. We do not evaluate equation~\ref{eq:gravpower} ourselves, but instead obtain $P_\Psi(k,z)$ directly from \texttt{CAMB} with the accuracy settings in Listing~\ref{list:camb}.

To calculate the lensing power spectrum from the binned primordial power spectrum, we note that the relationship between the primordial power spectrum and the linear matter power spectrum is 
\begin{equation}
            P_\mathrm{lin}(k,z) = 2\pi^2 k\mathcal{P}(k) G^2(z)\mathcal{T}^2(k)
            \label{eq:prim_to_lin}
\end{equation}
where $G^2(z)$ is a growth function and $\mathcal{T}^2(k)$ is a transfer function~\cite{hlozek2012atacama}.
There is also a transfer function between the linear and non-linear matter power spectrum per redshift and $k$ bin, i.e.
\begin{equation}
            P_m(k,z) = \mathcal{T}_\mathrm{non-lin}(k,z) P_\mathrm{lin}(k,z)
            \label{eq:lin_to_nonlin}
\end{equation}
We assume the transfer functions between the primordial power spectrum, $\mathcal{P}(k)$, the linear matter power spectrum, $P_\mathrm{lin}(k,z)$, and the non-linear matter power spectrum, $P_m(k,z)$, are all the same as in the $\Lambda$CDM model.  Thus, varying $\mathcal{P}(k)$ in a given $k$ bin gives us the following transfer function for $P_m(k,z)$:
\begin{equation}
    T^2(k,z) = \frac{P_m^{\mathrm{var}}(k,z)}{P_m^{\mathrm{fid}}(k,z)} = \frac{\mathcal{P^{\mathrm{var}}}(k)}{\mathcal{P}^{\mathrm{fid}}(k)} = \begin{cases} 
    1.05~\text{$\in$bin} \\
    1.00~\text{else}
    \end{cases}
    \label{eq:final_transfer}
\end{equation}
where ``var'' and ``fid'' indicate the varied and fiducial power spectra, respectively, and $\in$bin indicates for $k$-modes within the given $k$ bin.

We modify Eq~\ref{eq:lensing_int} by multiplying the integrand by the transfer function $T^2$ and calculate $C_\ell^{\kappa \kappa}$.  We then pass $C_\ell^{\kappa \kappa}$ to \texttt{CAMB}, which calculates the lensed power spectra from the unlensed power spectra we described above, following the procedure in~\cite{MacInnis:darkmatter}; specifically, we use the \texttt{hdPk}\footnote{\href{https://github.com/CMB-HD/hdPk}{https://github.com/CMB-HD/hdPk}} code developed for~\cite{MacInnis:darkmatter}. \\

With the lensed spectra in hand, we compute the Fisher matrix as described previously.  In addition to the primordial power spectrum in each $k$ bin, we vary the four $\Lambda\mathrm{CDM}$ parameters $\tau$, $\Omega_ch^2$, $\Omega_bh^2$, and $H_0$.  To compare with the $e^{-2\tau}\mathcal{P}(k)$ constraints described in Section~\ref{sec:method-current}, we use the python package \texttt{getdist}~\cite{Lewis:getdist} to add the $e^{-2\tau}\mathcal{P}(k)$ as derived parameters.
Specifically, we pass the Fisher matrix and the fiducial parameter values to \texttt{getdist} via the \texttt{getdist.gaussian\_mixtures.GaussianND} python class. From this, \texttt{getdist} generates Gaussian random samples of our varied parameter set, with the correct correlations between the parameters.  For each sample of the set of parameters, we construct $e^{-2\tau}\mathcal{P}(k)$ for each $k$ bin; these become our derived parameters.  We then use \texttt{getdist} to obtain the error on these derived parameters given the samples.

\section{Theory Models}
\label{sec:theory}

The pioneering inflationary scenarios of the late 1970s and early 1980s introduced a phase of accelerated expansion before the radiation-dominated hot Big Bang~\cite{Starobinsky:1980te,Guth:1980zm,Linde:1981mu,Albrecht:1982wi}.  Many inflationary and non-inflationary proposals for this primordial phase have since been developed~\cite{Afshordi:2025book}.  Inflation is usually driven by a scalar field, the inflaton, and a sufficiently long phase makes the observable Universe nearly homogeneous and spatially flat.  Quantum fluctuations generated during this phase can produce nearly scale-invariant scalar and tensor power spectra; hundreds of inflationary realizations and their predictions have been catalogued in~\cite{Martin:2013tda}.  Primordial tensor modes have not yet been detected, so here we focus on scalar modes, whose spectra can be measured precisely by CMB experiments such as SO and CMB-HD.  We use the following representative inflationary and non-inflationary models, which make distinct predictions for the scalar power spectrum.

\subsection{Inflationary Models}

Most simple inflationary models are characterized by a slow-roll inflation potential, $V(\phi)$, and the number of e-foldings, $N_*$, of expansion from the time observable modes exited the horizon to the end of inflation. The number of e-foldings is typically limited to be within $47 \lesssim N_* \lesssim 57$, with the upper limit coming from the highest scale of inflation currently allowed given the non-detection of tensor modes and instantaneous reheating, while the lower limit assumes a TeV-scale reheating temperature (e.g., \cite{CMB-S4:2016ple,Balkenhol:2025wms}). We will consider the following inflationary models: 

\begin{enumerate}
\item \textbf{Starobinsky inflation} adds an $R^2$ correction to the gravitational action; in its scalar-field representation, the scalaron rolls along an exponentially approached plateau~\cite{Starobinsky:1980te}.  \textbf{Higgs inflation} instead uses the Standard Model Higgs field with a non-minimal coupling to curvature~\cite{Bezrukov:2007ep}, while \textbf{exponential $\alpha$-attractors} obtain the same plateau behavior from a curved scalar-field geometry~\cite{Kallosh:2013yoa}.  Their potentials take the asymptotic form $V(\phi) \propto [1-\exp(-B\phi)]^2$, giving~\cite{Bianchi_2024}
\begin{align}
     & \alpha_s = -\frac{2}{N_*^2} + \mathcal{O}(N_*^{-3}), \\
     & n_s = 1 - \frac{2}{N_*} + \mathcal{O}(N_*^{-2}).
\end{align}
Higher-order terms change $n_s$ and $\alpha_s$ by less than $2.5\%$ over the $N_*$ range considered here.

\item \textbf{Polynomial $\alpha$-attractors} approach their inflationary plateau as an inverse power, $V(\phi) \simeq A-B\phi^{-\kappa}$, rather than exponentially~\cite{Kallosh:2022feu}.  This slower approach gives larger $n_s$ and $\alpha_s$:
\begin{align}
     & \alpha_s = -\frac{2}{N_*^2}\left(\frac{\kappa+1}{\kappa+2}\right) + \mathcal{O}(N_*^{-3}), \\
     & n_s = 1 - \frac{2}{N_*}\left(\frac{\kappa+1}{\kappa+2}\right) + \mathcal{O}(N_*^{-2}).
\end{align}
We consider the simplest case $\kappa=2$, as an example here.

\item \textbf{Quantum quadratic gravity} is an asymptotically free higher-derivative theory in which one-loop running generates slow-roll inflation toward the infrared~\cite{Liu:2025nwz}.  Its inflationary dynamics are well approximated by $V(\phi)=A-B\phi^{-1}$, yielding
\begin{align}
    & \alpha_s = -\frac{4}{3N_*^2} + \mathcal{O}(N_*^{-3}), \\
    & n_s = 1 - \frac{4}{3N_*} + \mathcal{O}(N_*^{-2}).
\end{align}

\item \textbf{Axion-monodromy inflation} uses string-theory effects to unwrap an axion's periodic field range, allowing a long field excursion and fractional-power monomial potentials, $V(\phi)\propto\phi^p$~\cite{Silverstein:2008sg,McAllister:2008hb}.  The bound $r<0.036$ limits $p\lesssim0.5$~\cite{BICEP:2021xfz}.  The $p=2/5$ construction of~\cite{Silverstein:2008sg}, for example, gives
\begin{align}
    & \alpha_s = -\frac{6}{5N_*^2} + \mathcal{O}(N_*^{-3}), \\
    & n_s = 1 - \frac{6}{5N_*} + \mathcal{O}(N_*^{-2}).
\end{align}

\item \textbf{Supersymmetric grand unified theories} can realize hybrid inflation: a gauge-singlet inflaton evolves along a nearly flat valley until grand-unified-theory fields trigger the end of inflation~\cite{Dvali:1994ms}.  Radiative corrections give the effective logarithmic potential $V(\phi)=A+B\ln\phi$.  For small $B$, corresponding to the $p\to0$ limit of monomial inflation,
\begin{align}
    & \alpha_s = -\frac{1}{N_*^2} + \mathcal{O}(N_*^{-3}), \\
    & n_s = 1 - \frac{1}{N_*} + \mathcal{O}(N_*^{-2}).
\end{align}
\end{enumerate}

\subsection{Non-inflationary Models}
\begin{enumerate}
\item \textbf{Renormalizable holographic cosmology} describes primordial fluctuations through stress-tensor correlators of a three-dimensional renormalizable quantum field theory dual to the four-dimensional cosmology~\cite{Coriano:2020zap,Afshordi:2022ive}.  Logarithmic renormalization-group running produces the correlated predictions
\begin{align}
    & \alpha_s = \frac{1}{[32.17+\ln(\theta_\star \ell_P)]^2}, \\
    & n_s = 1 - \frac{1}{32.17+\ln(\theta_\star \ell_P)}, \\
    & 0.1 < \theta_\star \ell_P < 10.
\end{align}

\item The \textbf{Bi-thermal Big Bang} is a bimetric scalar-tensor model in which matter and gravity have different propagation speeds.  Thermal fluctuations therefore begin inside the horizon without inflation, and the critical solution predicts~\cite{Afshordi:2016guo,Mylova:2021eld}
\begin{align}
    & \alpha_s = -1.8 \times 10^{-3}, \\
    & n_s = 0.96478 \pm 0.00064.
\end{align}

\item \textbf{Super-renormalizable holographic cosmology} replaces the earliest geometric description with a weakly coupled three-dimensional super-renormalizable quantum field theory~\cite{Afshordi:2016dvb,Afshordi:2017ihr}.  Its dimensionful effective coupling runs strongly with scale, producing a primordial spectrum that is not well approximated over the full observable range by the logarithmic expansion in Eq.~(\ref{eq:P(k)}). Instead, the power spectrum, at 2-loop order, is given by:
\begin{equation}
    \mathcal{P}(k) \simeq \frac{A_s}{1+\left(g k_0/k\right)\ln\left|k/(\beta g k_0)\right|+{\cal O}\left[\left(g k_0/k\right)^2\right]},\label{eq:HC}
\end{equation}
where $10^9 e^{-2\tau} A_s = 1.804 \pm 0.008$, $10^3g=-10.7 \pm 1.6$, and $\beta=2.7 \pm 0.3$ provide an excellent fit to the current CMB data (i.e., CMB-PAS constraints from this work shown in Table \ref{tab:pk-errors}). Nevertheless, the scalar spectrum becomes scale-invariant at large $k$, and thus can be easily distinguished from a red spectrum with logarithmic running using future CMB data. 
\end{enumerate}
We show the $n_s$ and $\alpha_s$ values or ranges of these models in Figures~\ref{fig:alpha_s-vs-n_s-data} and~\ref{fig:alpha_s-vs-n_s-forecasts}. The predictions for the super-renormalizable holographic cosmology (Eq.~\ref{eq:HC}) are compared to current constraints and forecasts in Figure~\ref{fig:primordialPk}.  

\section{Results}
\label{sec:results}

\begin{figure*}[t]
    \centering
    \begin{minipage}[t]{.49\textwidth}       \includegraphics[width=\textwidth]{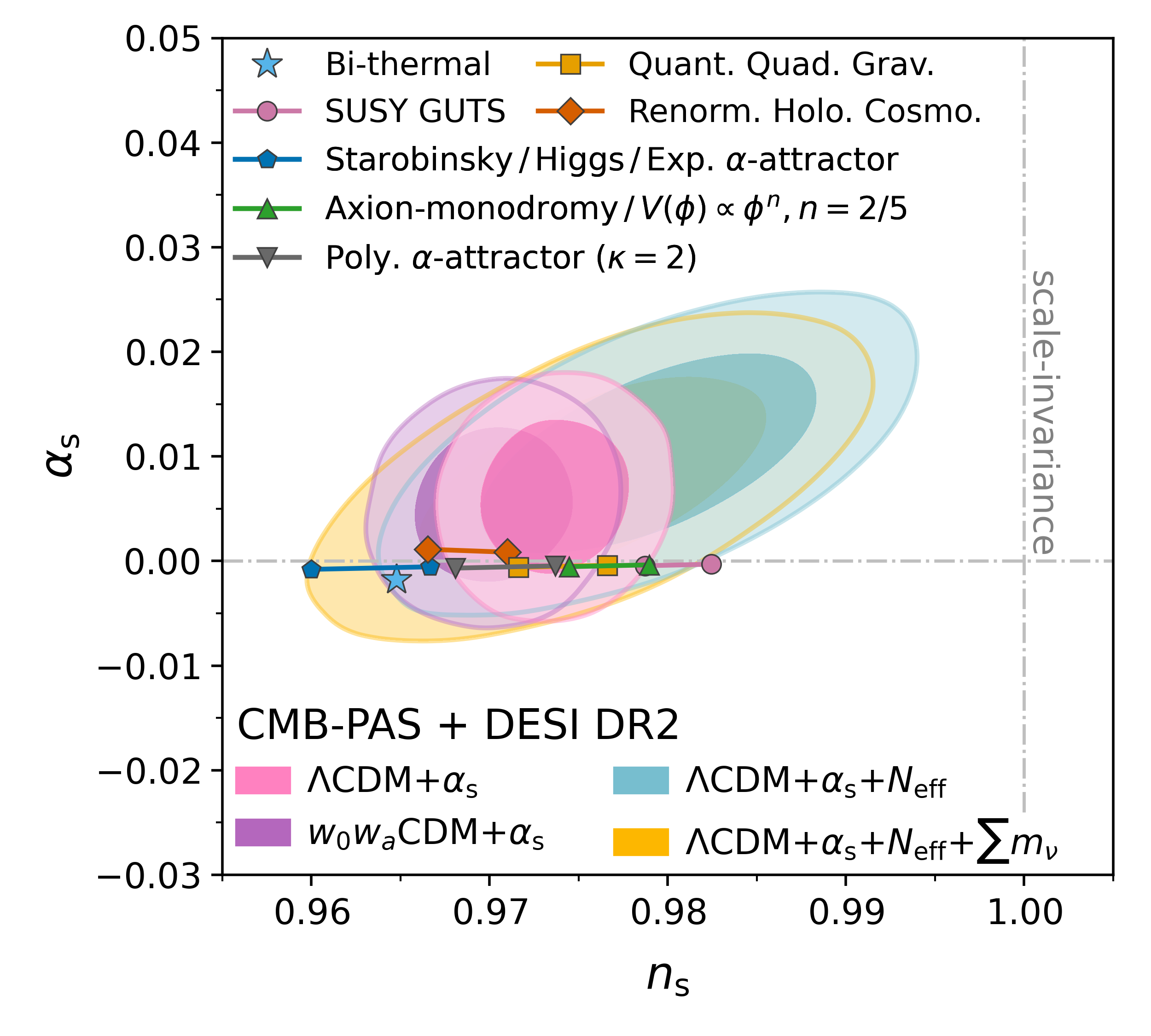}
        \caption{CMB-PAS plus DESI DR2 constraints on the scalar spectral index $n_s$ and its running $\alpha_s$ for four cosmological models: $\Lambda\mathrm{CDM}+\alpha_s$ (pink), $w_0w_a\mathrm{CDM}+\alpha_s$ (purple),  $\Lambda\mathrm{CDM}+\alpha_s+N_\mathrm{eff}$ (blue), and  $\Lambda\mathrm{CDM}+\alpha_s+N_\mathrm{eff}+\sum m_\nu$ (yellow).  The CMB-PAS likelihood combination includes lensed CMB $TT$, $TE$, $EE$ power spectra from \textit{Planck} 2018~\cite{Planck:2018vyg}, ACT DR6~\cite{AtacamaCosmologyTelescope:2025blo}, and SPT-3G D1~\cite{SPT-3G:2025bzu}, along with ACT-{\it{Planck}}-SPT (APS) CMB lensing~\cite{ACT:2025qjh}. The BAO data is from DESI DR2~\cite{DESI:2025zgx}.  Contours display the $1\sigma$ and $2\sigma$ confidence regions. Theory predictions from several models of inflation discussed in Section~\ref{sec:theory} are overlaid, spanning $N_* = 47$--$57$ e-folds where applicable. For the $\Lambda\mathrm{CDM}+\alpha_s$ model, the Bi-thermal, Starobinsky, and SUSY GUTS models are all disfavored by greater than $2\sigma$. The $w_0w_a\mathrm{CDM}+\alpha_s$ model shifts the mean values of $n_s$ and $\alpha_s$ to bring the Bi-thermal and Starobinsky models back within the $2\sigma$ contour. When $N_\mathrm{eff}$ and $\sum m_\nu$ are also varied, the contours broaden significantly due to parameter degeneracies; this yields \num{$n_s = 0.9758 \pm 0.0064$} and \num{$\alpha_s = 0.0080 \pm 0.0064$} and brings all models shown into consistency with the data.}
        \label{fig:alpha_s-vs-n_s-data}
    \end{minipage}
    \hfill
    \begin{minipage}[t]{.49\textwidth}        \includegraphics[width=\textwidth]{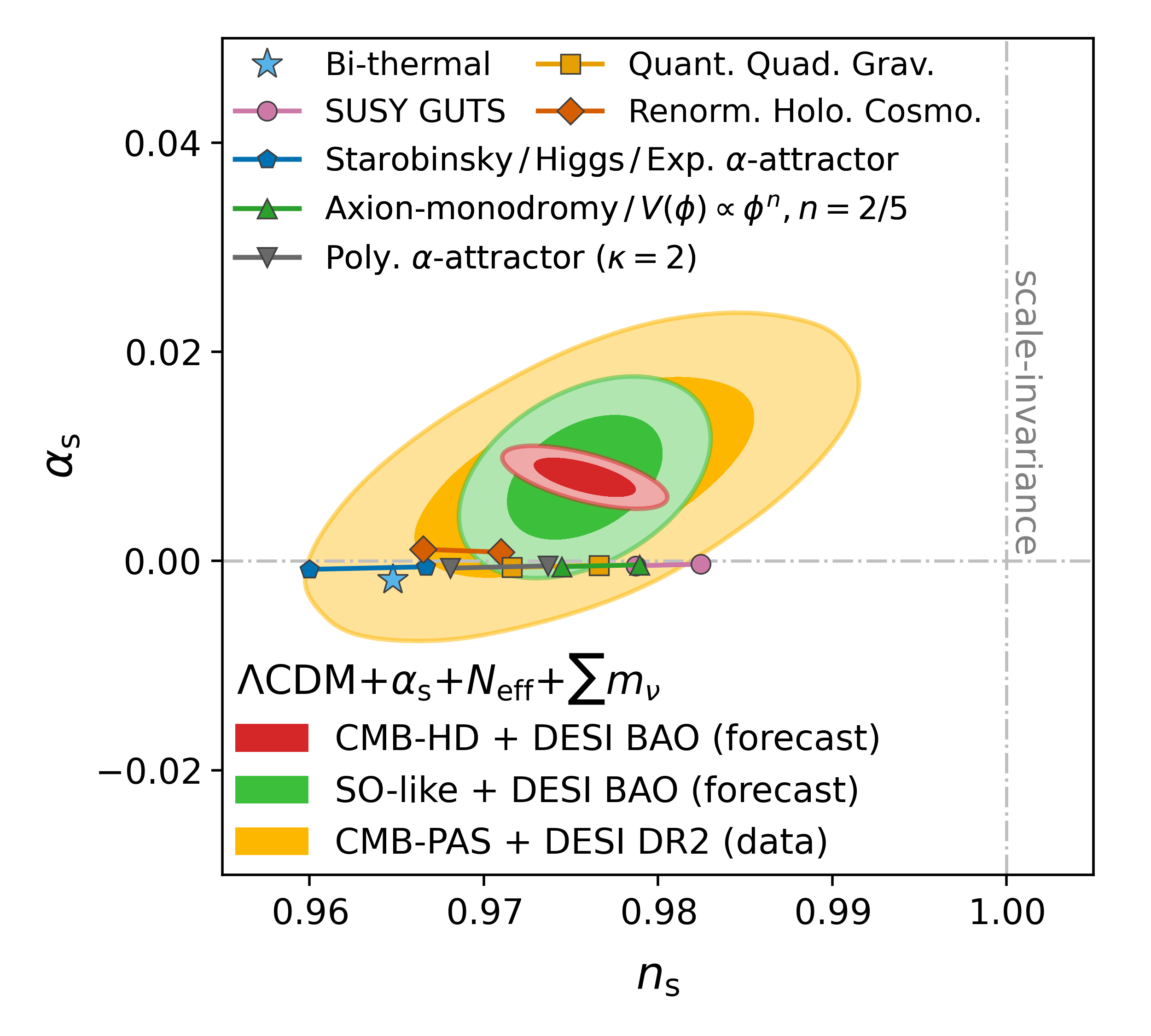}
         \caption{Current CMB-PAS plus DESI DR2 constraints on the scalar spectral index $n_s$ and its running $\alpha_s$ in a $\Lambda\mathrm{CDM}+\alpha_s+N_\mathrm{eff}+\sum m_\nu$ model are shown by the yellow contours (same contours as in Figure~\ref{fig:alpha_s-vs-n_s-data}).  These constraints are compared to projected $1\sigma$ and $2\sigma$ constraints from SO-like~\cite{SimonsObservatory:2025wwn} (green) and CMB-HD-like~\cite{CMB-HD:2022bsz} (red) experiments, combined with mock DESI BAO data~\cite{DESI:2016fyo}. The SO and CMB-HD forecasts are centered on the CMB-PAS plus DESI DR2 marginalized mean values of $n_\mathrm{s}$ and $\alpha_\mathrm{s}$. 
         Theory predictions from several models of inflation discussed in Section~\ref{sec:theory} are overlaid, spanning $N_* = 47$--$57$ e-folds where applicable. 
         While none of these inflation models are ruled out by the current CMB and BAO data for this cosmological parameter model, the forecasts show that SO and CMB-HD could rule out some or all of these inflation models by greater than $3\sigma$, assuming the best-fit values of $n_\mathrm{s}$ and $\alpha_\mathrm{s}$ remain as indicated by current data.} 
         \label{fig:alpha_s-vs-n_s-forecasts}
    \end{minipage}
\end{figure*}

Here, we present the cosmological parameter constraints from current data (Section~\ref{subsec:results-current}) and projected constraints from future data (Section~\ref{subsec:results-projected}).

\begin{table*}[t]
    \centering
    \begin{tabular}{l@{\hskip 1.5em} c c@{\hskip 1em} c@{\hskip 1em} c@{\hskip 1em} c} 
    \toprule
    \toprule
     & P-ACT-LB & \multicolumn{4}{c}{CMB-PAS+DESI DR2} \\ 
    \cmidrule(){2-2} \cmidrule(lr){3-6}
    Parameter  & $\Lambda\mathrm{CDM}+\alpha_s$  & $\Lambda\mathrm{CDM}+\alpha_s$ & $w_0w_a\mathrm{CDM}+\alpha_s$ & $\Lambda\mathrm{CDM}+\alpha_s + N_{\mathrm{eff}}$ &  $\Lambda\mathrm{CDM}+\alpha_s+N_{\mathrm{eff}}+\sum m_\nu$ \\
    \midrule
    $\Omega_\mathrm{b} h^2$ \dotfill & 0.02249 $\pm$ 0.00012 & 0.02240 $\pm$ 0.00010 & 0.02236 $\pm$ 0.00010 & 0.02248 $\pm$ 0.00013 & 0.02244 $\pm$ 0.00013 \\
    $\Omega_\mathrm{c} h^2$ \dotfill & 0.1179 $\pm$ 0.0008 & 0.1181 $\pm$ 0.0006 & 0.1196 $\pm$ 0.0007 & 0.1203 $\pm$ 0.0023 & 0.1194 $\pm$ 0.0023 \\
    $100 \theta_\mathrm{MC}$ \dotfill & 1.04087 $\pm$ 0.00025 & 1.04090 $\pm$ 0.00023 & 1.04073 $\pm$ 0.00023 & 1.04072 $\pm$ 0.00029 & 1.04080 $\pm$ 0.00030 \\
    $\tau$ \dotfill & 0.0611 $\pm$ 0.0062 & 0.0611 $\pm$ 0.0051 & 0.0537 $\pm$ 0.0054 & 0.0601 $\pm$ 0.0052 & 0.0585 $\pm$ 0.0051 \\
    $\ln(10^{10} A_\mathrm{s})$ \dotfill & 3.055 $\pm$ 0.012 & 3.053 $\pm$ 0.010 & 3.040 $\pm$ 0.010 & 3.055 $\pm$ 0.010 & 3.050 $\pm$ 0.010 \\
    $n_\mathrm{s}$ \dotfill & 0.9741 $\pm$ 0.0033 & 0.9737 $\pm$ 0.0027 & 0.9702 $\pm$ 0.0029 & 0.9790 $\pm$ 0.0062 & 0.9758 $\pm$ 0.0064 \\
    $\alpha_\mathrm{s}$ \dotfill & 0.0062 $\pm$ 0.0053 & 0.0062 $\pm$ 0.0049 & 0.0054 $\pm$ 0.0049 & 0.0102 $\pm$ 0.0063 & 0.0080 $\pm$ 0.0064 \\
    $w_0$ \dotfill & --- & --- & -0.433 $\pm$ 0.202 & --- & --- \\
    $w_a$ \dotfill & --- & --- & -1.69 $\pm$ 0.56 & --- & --- \\
    $N_\mathrm{eff}$ \dotfill & --- & --- & --- & 3.18 $\pm$ 0.14 & 3.11 $\pm$ 0.14 \\
    $\sum m_\nu$ [eV] \dotfill & --- & --- & --- & --- & $<$ 0.056 \\
    \midrule
    $H_0$ \dotfill & 68.16 $\pm$ 0.36 & 68.10 $\pm$ 0.24 & 63.81 $\pm$ 1.85 & 68.91 $\pm$ 0.88 & 68.73 $\pm$ 0.87 \\
    $\sigma_8$ \dotfill & 0.8120 $\pm$ 0.0045 & 0.8142 $\pm$ 0.0039 & 0.7858 $\pm$ 0.0164 & 0.8206 $\pm$ 0.0077 & 0.8241 $\pm$ 0.0078 \\

    \bottomrule
    \end{tabular}
    \caption{Cosmological parameter marginalized means and $1\sigma$ uncertainties from current data. The leftmost column 
    shows constraints from P-ACT-LB~\cite{AtacamaCosmologyTelescope:2025nti}; P-ACT-LB includes lensed CMB $TT$, $TE$, $EE$ power spectra from \textit{Planck} 2018~\cite{Planck:2019nip} and ACT DR6~\cite{AtacamaCosmologyTelescope:2025blo}, the joint {\it{Planck}} and ACT CMB lensing spectrum presented in~\cite{ACT:2023dou, ACT:2023kun}, and DESI DR1 BAO data~\cite{DESI:DR1}. The remaining columns use the CMB-PAS plus DESI DR2 data combination described in 
    Section~\ref{sec:current-data}, which adds the SPT-3G D1 CMB spectra~\cite{SPT-3G:2025bzu}
    and replaces the CMB lensing spectrum with that from a joint ACT-\textit{Planck}-SPT (APS) lensing analysis~\cite{ACT:2025qjh}.  Parameters marked ``---'' are held fixed at their $\Lambda\mathrm{CDM}$ values. The Hubble parameter $H_0$ is given in units of km\,s$^{-1}$\,Mpc$^{-1}$ and is a derived parameter in all runs using current data. 
    $\sum m_\nu$ is reported as a $95\%$ upper limit. All chains have Gelman-Rubin statistic $R-1 < 0.01$ after removal of a 50\% burn-in.  In the $w_0 w_a$CDM+$\alpha_\mathrm{s}$ model, we find $n_s$ is shifted to lower values than in the $\Lambda\mathrm{CDM}+\alpha_s$ model; we also still observe values of $w_0$ and $w_a$ that disfavor a cosmological constant by greater than $2\sigma$.  We find that the constraints on $n_s$ and $\alpha_s$ broaden by a factor of about 2 and 1.5, respectively, when $N_\mathrm{eff}$ and $\sum m_\nu$ are also allowed to vary; this brings models such as Starobinsky inflation, SUSY GUTS, and the Bi-thermal Big Bang model, which were disfavored by the $\Lambda\mathrm{CDM} + \alpha_s$ model, back into agreement with the data, as also shown in Figure~\ref{fig:alpha_s-vs-n_s-data}. We note that $\sum m_\nu < 0.056$ at 95\% confidence level even when $N_\mathrm{eff}$ and $\alpha_s$ are also varied in addition to the $\Lambda$CDM parameters; this remains in mild tension with neutrino oscillation experiments.}
    \label{tab:current-constraints}
\end{table*}

\begin{figure*}[t]
    \centering  \includegraphics[width=0.8\linewidth]{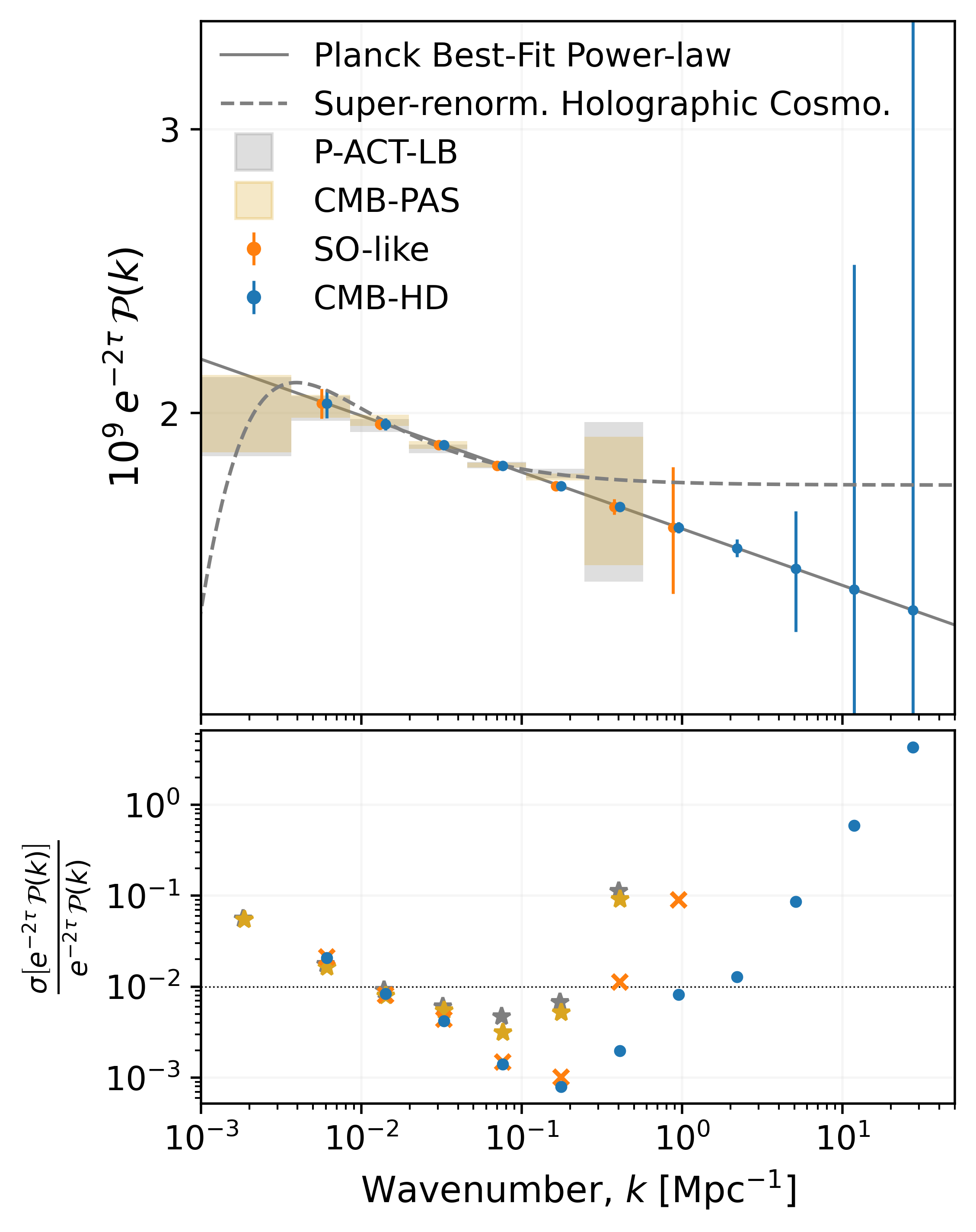}
    \caption{ {\it{Top:}} Current and projected constraints on $e^{-2\tau}\mathcal{P}(k)$ for a binned primordial power spectrum. The four $\Lambda\mathrm{CDM}$ parameters $\theta$, $\tau$, $\Omega_c h^2$, and $\Omega_b h^2$ are varied here as well as $e^{-2\tau}\mathcal{P}(k)$ for each $k$ bin.  We show the marginalized mean values and $1\sigma$ errors on $e^{-2\tau}\mathcal{P}(k)$ for each $k$ bin for {\it{Planck}}, ACT, and DESI DR1 data (P-ACT-LB)~\cite{AtacamaCosmologyTelescope:2025nti} and {\it{Planck}}, ACT, and SPT data (CMB-PAS)~\cite{SPT-3G:2025bzu} in grey and gold, respectively. P-ACT-LB includes the joint {\it{Planck}} 2018 and ACT DR6 CMB lensing from~\cite{ACT:2023kun}, while CMB-PAS includes the ACT, {\it{Planck}}, and SPT (APS) CMB lensing (extended version) from~\cite{ACT:2025qjh}. The $k$ bin centers and prior ranges are listed in Table~\ref{tab:P(k)}; we also show these constraints with finer binning in Appendix~\ref{sec:AppendixPk}. The inclusion of SPT data and the improved CMB lensing data likely accounts for the improvement of CMB-PAS over P-ACT-LB.  The inclusion of DESI DR2 BAO data minimally impacts the CMB-PAS constraints, so we do not include it here (see Appendix~\ref{sec:AppendixPk}).  We find excellent agreement with the best-fit power-law model from Planck 2018 data~\cite{Planck:2018vyg} (grey solid line).
    We also show projected constraints on $e^{-2\tau}\mathcal{P}(k)$ for SO-like~\cite{SimonsObservatory:2025wwn} (orange) and CMB-HD-like~\cite{CMB-HD:2022bsz} (blue) experiments, and the theory expectation for an illustrative non-power-law model given by Eq.~\ref{eq:HC} (dashed grey).  
    {\it{Bottom:}} The fractional uncertainty on $e^{-2\tau}\mathcal{P}(k)$ for each $k$ bin.  We find, for example, that CMB-PAS constrains $e^{-2\tau}\mathcal{P}(k = \mathrm{0.2~Mpc^{-1}})$ to \num{0.5\%}, and SO-like and CMB-HD-like surveys would constrain this to \num{0.1\%}. CMB-HD would also extend these measurements to smaller scales to constrain $e^{-2\tau}\mathcal{P}(k = \mathrm{30~Mpc^{-1}})$ to within a factor of \num{ten}. We find minimal change to the CMB-HD forecasts when also varying baryonic physics parameters, as shown in Figure~\ref{fig:HDPk-w-feedback}.}
    \label{fig:primordialPk}
\end{figure*}

\subsection{Current Constraints}
\label{subsec:results-current}

We show in Figure~\ref{fig:alpha_s-vs-n_s-data} the CMB-PAS plus DESI DR2 $1\sigma$ and $2\sigma$ constraints on the scalar spectral index $n_s$ and its running $\alpha_s$ for four cosmological models: $\Lambda\mathrm{CDM}+\alpha_s$ (pink), $w_0w_a\mathrm{CDM}+\alpha_s$ (purple),  $\Lambda\mathrm{CDM}+\alpha_s+N_\mathrm{eff}$ (blue), and $\Lambda\mathrm{CDM}+\alpha_s+N_\mathrm{eff}+\sum m_\nu$ (yellow).  We also list the marginalized mean cosmological parameter values and $1\sigma$ errors for these models in Table~\ref{tab:current-constraints}.  For comparison with previous data combinations, in Table~\ref{tab:current-constraints} we also show constraints for the $\Lambda\mathrm{CDM}+\alpha_s$ model from P-ACT-LB~\cite{AtacamaCosmologyTelescope:2025blo, AtacamaCosmologyTelescope:2025nti}; P-ACT-LB includes the latest {\it{Planck}} and ACT CMB spectra~\cite{AtacamaCosmologyTelescope:2025blo}, the joint {\it{Planck}} and ACT CMB lensing spectrum from~\cite{ACT:2023dou, ACT:2023kun}, and the DESI DR1 BAO data~\cite{DESI:DR1}.

We find that varying $\alpha_s$ in the $\Lambda\mathrm{CDM}+\alpha_s$ model does not change constraints on $n_s$ significantly compared to the $\Lambda\mathrm{CDM}$ model constraints presented in~\cite{AtacamaCosmologyTelescope:2025blo, SPT-3G:2025bzu,McDonough:2025lzo}. Our results are also consistent with previous findings that have pointed out a slight preference for positive running from various combinations of CMB and BAO data~\cite{AtacamaCosmologyTelescope:2025nti, Fairbairn:2025fko,Garny:2026gcs}.  
The $w_0w_a\mathrm{CDM} + \alpha_s$ model yields similar error contours to the $\Lambda\mathrm{CDM}+\alpha_s$ model; however, the central value of $n_s$ is shifted to lower values, as also noted in~\cite{Chudaykin:2026amr}.  We also still see the $2\sigma$ shift away from $\Lambda$CDM $w$ values, as pointed out in~\cite{DESI:2025zgx}.

We also find that additionally varying $N_\mathrm{eff}$ and, to a lesser extent, $\sum m_\nu$ broadens the $n_s$ and $\alpha_s$ contours by about a factor of 2 and 1.5, respectively, compared to the $\Lambda\mathrm{CDM}+\alpha_s$ model.  This is due to the degeneracy between the parameters, in particular between $N_\mathrm{eff}$ and $n_s$.  Varying $\sum m_\nu$ in the $\Lambda\mathrm{CDM}+\alpha_s+N_\mathrm{eff}+\sum m_\nu$ model also shifts $n_s$ error contours to lower values compared to the $\Lambda\mathrm{CDM}+\alpha_s+N_\mathrm{eff}$ model, allowing popular inflation models, such as Starobinsky inflation, to be fully consistent with the data.  

We also show in Figures~\ref{fig:alpha_s-vs-n_s-data} and~\ref{fig:alpha_s-vs-n_s-forecasts} the predictions for $n_\mathrm{s}$ and $\alpha_\mathrm{s}$ from various early Universe models discussed in Section~\ref{sec:theory}.  The light blue star is the prediction made by the Bi-Thermal Big Bang model.  The inflation models shown are Starobinsky/Higgs/Exponential $\alpha$-attractor (dark blue pentagons), Polynomial $\alpha$-attractor for $\kappa=2$ (grey triangles), Axion-monodromy/Monomial potential for $p=2/5$ (green triangles), Quantum Quadratic Gravity (orange squares), and Spontaneously Broken Supersymmetric Grand Unified Theories (pink circles). All inflation models are shown for $N_\star$ e-folds with $N_\star$ ranging from 47 to 57.  The predictions of Renormalizable Holographic Cosmologies (red diamonds) span the range $0.1 < \theta_\star \ell_P < 10$.  Each of these models predicts a value of $\alpha_s$ that is close to, but not exactly zero, while making predictions for $n_s$ in the range of 0.96 to 0.983.

In Figure~\ref{fig:primordialPk} (top panel), we show the marginalized means and $1\sigma$ error bars on $e^{-2\tau}\mathcal{P}(k)$ for a binned primordial power spectrum from the P-ACT-LB (grey) and CMB-PAS (gold) datasets.  Here we vary the four $\Lambda\mathrm{CDM}$ parameters $\theta$, $\tau$, $\Omega_c h^2$, and $\Omega_b h^2$ as well as $e^{-2\tau}\mathcal{P}(k)$ for each $k$ bin.  We use the bin centers and priors given in Table~\ref{tab:P(k)}.  Both P-ACT-LB and CMB-PAS results are consistent with a primordial power spectrum given by a power law, as also found in~\cite{AtacamaCosmologyTelescope:2025nti,Chandra:2026byw}. In particular, both are consistent with the fiducial power spectrum from Equation~\ref{eq:P(k)} using the cosmological parameter values listed in Table~\ref{tab:fisher} from the Planck 2018~\cite{Planck:2018vyg} marginalized mean values (grey theory line).

In the bottom panel of Figure~\ref{fig:primordialPk}, we show the fractional uncertainty on $e^{-2\tau}\mathcal{P}(k)$ for each $k$ bin.  We find that the CMB-PAS constraints improve over the P-ACT-LB constraints for some of the parameter range.  For example, at $k\approx0.2~\mathrm{Mpc}^{-1}$, P-ACT-LB and CMB-PAS constrain $e^{-2\tau}\mathcal{P}(k)$ to 0.7\% and 0.5\%, respectively.  This is likely due to the addition of SPT-3G data, as well as the APS CMB lensing in the CMB-PAS dataset.  We note that these constraints depend on the binning and show constraints with the finer binning used in~\cite{AtacamaCosmologyTelescope:2025nti} in Figure~\ref{fig:30bin} in Appendix~\ref{sec:AppendixPk}.
Specifically, for 30 $k$ bins, instead of seven, we find 
at $k\approx0.2~\mathrm{Mpc}^{-1}$ that P-ACT-LB and CMB-PAS constrain $e^{-2\tau}\mathcal{P}(k)$ to 2.1\% and 1.6\%, respectively.  The SO forecast in~\cite{SimonsObservatory:2018koc}, assuming 20 $k$ bins, projects a constraint of about 0.5\% for $k\approx0.2~\mathrm{Mpc}^{-1}$; in this work, we forecast a constraint of 0.1\% for $k\approx0.2~\mathrm{Mpc}^{-1}$ using seven wider bins.  We also find that adding DESI DR2 BAO data to CMB-PAS does not significantly impact the error bars, as we show in Figure~\ref{fig:comparePk7bin} in Appendix~\ref{sec:AppendixPk}. We provide the marginalized means and $1\sigma$ error bars for $e^{-2\tau}\mathcal{P}(k)$ for P-ACT-LB and CMB-PAS in Table~\ref{tab:pk-errors} in Appendix~\ref{sec:AppendixPk}.

\begin{table*}[t]
    \centering
    \begin{tabular}{l@{\hskip 1em} c@{\hskip 1em} c@{\hskip 1em} c@{\hskip 1em} c@{\hskip 1em} c@{\hskip 1em} c@{\hskip 1em} c@{\hskip 1em} c@{\hskip 1em}} 
    \toprule
    \toprule

    & \multicolumn{2}{c}{$\Lambda\mathrm{CDM}+\alpha_s$} & \multicolumn{2}{c}{$w_0w_a\mathrm{CDM}+\alpha_s$} & \multicolumn{2}{c}{$\Lambda\mathrm{CDM}+\alpha_s + N_{\mathrm{eff}}$} & \multicolumn{2}{c}{$\Lambda\mathrm{CDM}+\alpha_s+N_{\mathrm{eff}}+\sum m_\nu$} \\
    \cmidrule(lr){2-3} \cmidrule(lr){4-5} \cmidrule(lr){6-7} \cmidrule(lr){8-9}
    Parameter & SO + DESI & HD + DESI & SO + DESI & HD + DESI & SO + DESI & HD + DESI & SO + DESI & HD + DESI \\
    \midrule
   $\Omega_\mathrm{b} h^2$ \dotfill & $0.000036$ & $0.000016$ & $0.000036$ & $0.000018$ & $0.000047$ & $0.000026$ & $0.000048$ & $0.000026$\\
   $\Omega_\mathrm{c} h^2$ \dotfill & $0.00037$ & $0.00035$ & $0.00039$ & $0.00036$ & $0.00082$ & $0.00039$ & $0.00087$ & $0.00039$\\
   $H_0$ \dotfill & $0.14$ & $0.13$ & $1.6$ & $1.2$ & $0.38$ & $0.18$ & $0.41$ & $0.31$\\
   $\tau$ \dotfill & $0.0034$ & $0.0033$ & $0.0043$ & $0.0042$ & $0.0035$ & $0.0033$ & $0.0046$ & $0.0046$\\
   $\ln(10^{10} A_\mathrm{s})$ & $0.0060$ & $0.0058$ & $0.0078$ & $0.0075$ & $0.0060$ & $0.0058$ & $0.0089$ & $0.0085$\\
   $n_\mathrm{s}$ \dotfill & $0.0018$ & $0.0017$ & $0.0019$ & $0.0017$ & $0.0028$ & $0.0019$ & $0.0030$ & $0.0019$\\
   $\alpha_\mathrm{s}$ \dotfill & $0.0026$ & $0.0011$ & $0.0026$ & $0.0015$ & $0.0038$ & $0.0012$ & $0.0039$ & $0.0012$\\
   $w_0$ \dotfill & --- & --- & $0.133$ & $0.107$ & --- & --- & --- & ---\\
   $w_a$ \dotfill & --- & --- & $0.31$ & $0.26$ & --- & --- & --- & ---\\
   $N_\mathrm{eff}$ \dotfill & --- & --- & --- & --- & $0.052$ & $0.015$ & $0.056$ & $0.015$\\
   $\sum m_\nu$ [eV] \dotfill & --- & --- & --- & --- & --- & --- & $0.031$ & $0.028$\\
    \bottomrule
    \end{tabular}
    \caption{Projected $1\sigma$ marginalized parameter uncertainties from Fisher estimates for SO-like~\cite{SimonsObservatory:2025wwn} and CMB-HD-like~\cite{CMB-HD:2022bsz} experiments, combined with mock DESI BAO data~\cite{DESI:2016fyo}.  This is shown for four cosmological parameter models: $\Lambda\mathrm{CDM}+\alpha_s$, $w_0w_a\mathrm{CDM}+\alpha_s$,  $\Lambda\mathrm{CDM}+\alpha_s+N_\mathrm{eff}$, and  $\Lambda\mathrm{CDM}+\alpha_s+N_\mathrm{eff}+\sum m_\nu$.  
    The Fisher matrix estimates include a Gaussian prior of $\sigma(\tau) = 0.005$ on the optical depth (see Table~\ref{tab:fisher}). Parameters marked ``---'' are held fixed to standard values. $H_0$ is given in km\,s$^{-1}$\,Mpc$^{-1}$ and $\sum m_\nu$ in eV. We find that SO and CMB-HD can significantly reduce the degeneracy between $n_s$ and $N_\mathrm{eff}$ that exists with current data (see also Figures~\ref{fig:alpha_s-vs-n_s-forecasts} and~\ref{fig:hd_so_pas_triangle}). We also find that CMB-HD can significantly improve constraints on $H_0$, $\Omega_b h^2$, $\Omega_c h^2$, $n_s$, $N_\mathrm{eff}$, and $\alpha_s$ compared to SO (see also Figure~\ref{fig:hd_so_pas_triangle}). Fisher forecasts shown are computed with the \texttt{CLASS} code; we show agreement with \texttt{CAMB} Fisher forecasts, as well as \texttt{CAMB} and \texttt{CLASS} MCMC runs in Table~\ref{tab:camb_class_errs} and Figure~\ref{fig:camb_class_triangle} of Appendix~\ref{sec:stabillity}.} 
    \label{tab:projected-constraints}
\end{table*}

\begin{figure*}[t]
    \centering
    \includegraphics[width=\linewidth]{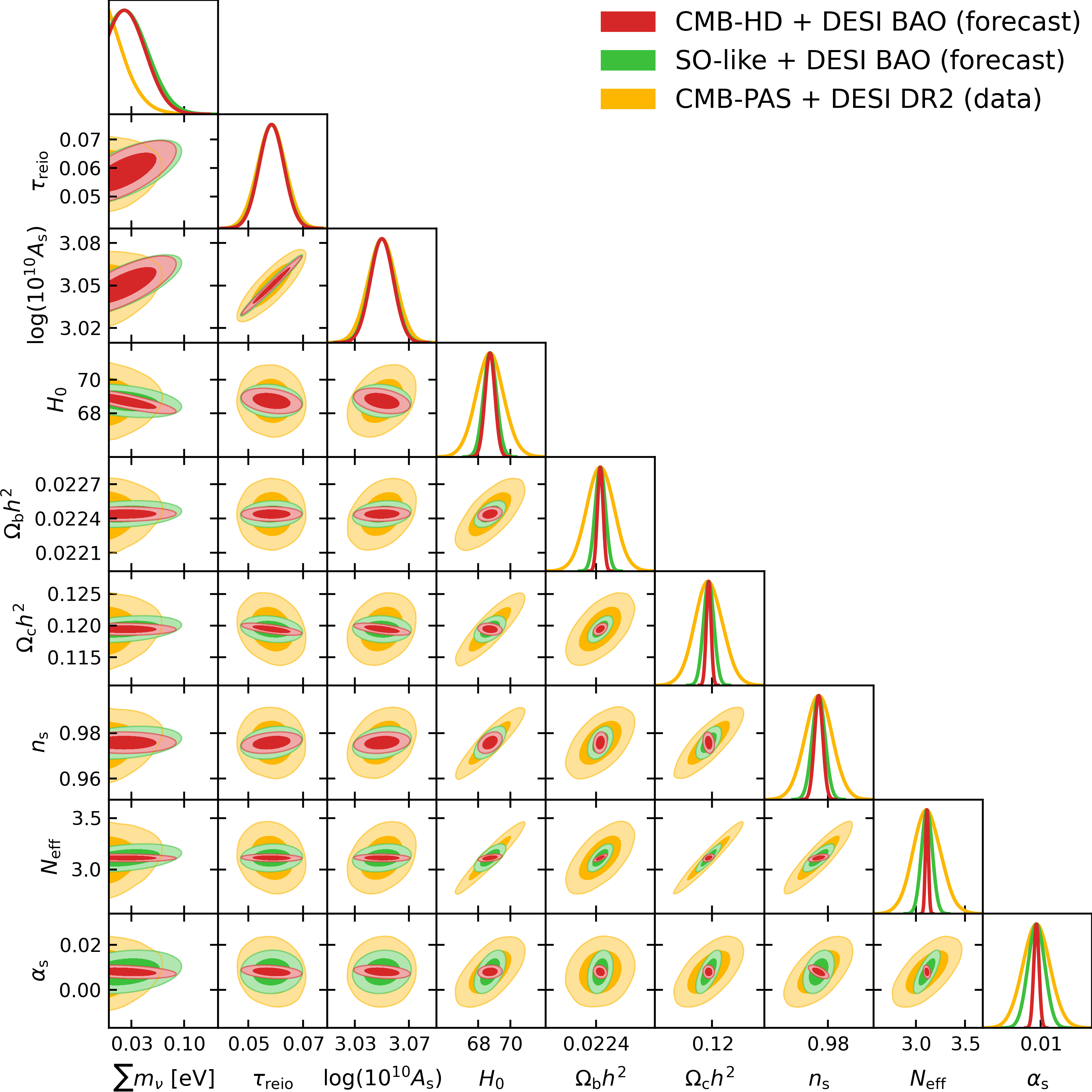}
    \caption{Current and projected cosmological parameter constraints from the combination of CMB, CMB lensing, and BAO data in a $\Lambda\mathrm{CDM}+\alpha_s+N_\mathrm{eff}+\sum m_\nu$ model. The current constraints (yellow contours) are obtained from running MCMC chains on CMB and CMB lensing data from {\it{Planck}}, ACT, and SPT (CMB-PAS) combined with DESI DR2 data (see Sections~\ref{sec:current-data} and~\ref{sec:method-current} for details on datasets and methods, respectively.) The projected constraints are obtained from a Fisher analysis of mock CMB and CMB lensing data from either an SO-like (green) or CMB-HD-like (red) experiment, combined with mock DESI BAO data (see Sections~\ref{sec:future-data} and~\ref{sec:method-projected} for details on mock datasets and methods, respectively.)
    Contours display the $1\sigma$ and $2\sigma$ confidence regions. Projected constraints for SO and CMB-HD are centered on the marginalized mean parameter values from the CMB-PAS plus DESI DR2 data. Parameters are ordered left-to-right roughly by degree of improvement. In particular, we find significant improvement in constraints on $H_0$, $\Omega_b h^2$, $\Omega_c h^2$, $n_s$, $N_\mathrm{eff}$, and $\alpha_s$, with SO plus DESI compared to current CMB-PAS plus DESI DR2 data, and with CMB-HD compared to SO.}  
    \label{fig:hd_so_pas_triangle}
\end{figure*}

\subsection{Projected Constraints}
\label{subsec:results-projected}

Figure~\ref{fig:alpha_s-vs-n_s-forecasts} shows the current and projected constraints on $n_\mathrm{s}$ and $\alpha_\mathrm{s}$ in a $\Lambda\mathrm{CDM}+\alpha_s+N_\mathrm{eff}+\sum m_\nu$ model, along with predictions for $n_\mathrm{s}$ and $\alpha_\mathrm{s}$ from the early Universe models discussed in Section~\ref{sec:theory}.  The current parameter constraints, shown as the yellow contour in both Figures~\ref{fig:alpha_s-vs-n_s-data} and~\ref{fig:alpha_s-vs-n_s-forecasts}, are obtained from the CMB-PAS plus DESI DR2 datasets described in Section~\ref{sec:current-data}.  The projected constraints for an SO-like and CMB-HD-like experiment combined with mock DESI BAO data are shown as the green and red contours, respectively; we center these contours on the marginalized mean values of $n_\mathrm{s}$ and $\alpha_\mathrm{s}$ from the CMB-PAS plus DESI DR2 data. As also shown in Figure~\ref{fig:alpha_s-vs-n_s-data}, Figure~\ref{fig:alpha_s-vs-n_s-forecasts} displays the theoretical models described in section~\ref{sec:theory} as the lines and points on the plot.  

While the CMB-PAS plus DESI DR2 data do not rule out any early Universe model considered here in the context of this nine-parameter cosmological model, we see that SO-like and CMB-HD-like experiments can reduce parameter degeneracies and narrow the range of viable models. To determine how well each dataset can constrain a given early Universe model, we calculate the Mahalanobis distance~\cite{mahalanobis}, given by
\begin{equation} \label{eq:Mahalanobis}
        d_M = \sqrt{(\vec{x} - \vec{u})^TC^{-1}(\vec{x}-\vec{u}),}
\end{equation}
where $C^{-1}$ is the inverse covariance matrix for the parameters (given by the Fisher matrix), $\vec{u}$ is a vector of the measured parameter values, and $\vec{x}$ is the model prediction for each parameter (in this case, $n_\mathrm{s}$ and $\alpha_\mathrm{s}$).  For the vector $\vec{u}$, we use the marginalized mean values of $\alpha_\mathrm{s}$ and $n_\mathrm{s}$ from the CMB-PAS plus DESI DR2 constraints in the $\Lambda$CDM+$\alpha_\mathrm{s}$+$N_\mathrm{eff}$+$\sum m_\nu$ model.  If we assume that the values of $n_\mathrm{s}$ and $\alpha_\mathrm{s}$ remain as indicated by current data, SO would rule out the Bi-thermal Big Bang model, SUSY GUTS, and Starobinsky inflation at the $3\sigma$ level.  CMB-HD would rule out all early Universe models considered by at least $6\sigma$ (SUSY GUTS) and by as much as $17\sigma$ (Starobinsky). 

In Table~\ref{tab:projected-constraints}, we show the $1\sigma$ marginalized parameter uncertainties from Fisher estimates for both SO-like and CMB-HD-like experiments, combined with mock DESI BAO data, for the four cosmological parameter models considered. We find that a CMB-HD-like experiment can maintain an uncertainty on $N_\mathrm{eff}$ of $\sigma(N_\mathrm{eff})\approx 0.015$, even when varying $\alpha_s$.  Figure~\ref{fig:hd_so_pas_triangle} shows the degeneracy between the cosmological parameters and, in particular, between $N_\mathrm{eff}$ and $n_s$.  Figure~\ref{fig:hd_so_pas_triangle} also shows the latest cosmological parameter constraints from the CMB-PAS plus DESI DR2 data (yellow), as well as projected constraints from SO-like (green) and CMB-HD-like (red) experiments plus mock DESI BAO data.  The contours show the $1\sigma$ and $2\sigma$ confidence regions, and the projected constraints for SO and CMB-HD are centered on the marginalized mean parameter values from the CMB-PAS plus DESI DR2 data. We find improvement in constraints on $H_0$, $\Omega_b h^2$, $\Omega_c h^2$, $n_s$, $N_\mathrm{eff}$, and $\alpha_s$, with SO plus DESI compared to the current CMB-PAS plus DESI DR2 data, and also with CMB-HD compared to SO.

The Fisher forecasts shown in Table~\ref{tab:projected-constraints} and Figure~\ref{fig:hd_so_pas_triangle} are computed with the \texttt{CLASS} code using the accuracy settings discussed in Appendix~\ref{sec:accuracy}.  Using the \texttt{CAMB} accuracy settings, also discussed in Appendix~\ref{sec:accuracy}, we find good agreement between \texttt{CAMB} and \texttt{CLASS} Fisher forecasts (to within \num{5\%}), as well as agreement between \texttt{CAMB} and \texttt{CLASS} MCMC runs; \texttt{CAMB} Fisher and MCMC errors agree to within \num{7\%}, and \texttt{CLASS} Fisher and MCMC errors agree to within \num{4\%}. We show this agreement in Table~\ref{tab:camb_class_errs} and Figure~\ref{fig:camb_class_triangle} of Appendix~\ref{sec:stabillity}. \\

We show in Figure~\ref{fig:primordialPk} the projected constraints on $e^{-2\tau}\mathcal{P}(k)$ for SO-like (orange) and CMB-HD-like (blue) experiments. These results are presented for a model that varies the four $\Lambda\mathrm{CDM}$ parameters $\tau$, $\Omega_ch^2$, $\Omega_bh^2$, and $H_0$, as well as the values of $e^{-2\tau}\mathcal{P}(k)$ in the $k$ bins given in Table~\ref{tab:P(k)}.  We described the method to obtain these constraints in Section~\ref{sec:method-projected-Pk}.  These forecasts can be compared to current constraints from P-ACT-LB (in grey) and CMB-PAS (in gold).  The top panel gives the error bars centered on the fiducial model from Eq.~\ref{eq:P(k)} using the fiducial cosmological parameters in Table~\ref{tab:fisher}, and the bottom panel shows the fractional error with respect to that fiducial model.  We see that for $k > 0.07 \ \mathrm{Mpc}^{-1}$, both SO and CMB-HD will improve upon current constraints, and in particular, both will achieve constraints on $e^{-2\tau}\mathcal{P}(k = \mathrm{0.2~Mpc^{-1}})$ of about \num{0.1\%}, a factor of five improvement over current uncertainties.  CMB-HD will constrain $e^{-2\tau}\mathcal{P}(k)$ to better than \num{2\%} for $k < \mathrm{2~Mpc^{-1}}$, and will extend these measurements to $k = \mathrm{30~Mpc^{-1}}$ with uncertainties within a factor of \num{ten}, assuming the fiducial model.  We find that the inclusion of mock DESI BAO data has minimal impact, as is also the case for current data (see Figure~\ref{fig:comparePk7bin}); thus, we do not include it here. We provide the current and projected constraints on the binned primordial power spectrum in Table~\ref{tab:pk-errors}. 

\subsubsection{Forecasts When Varying Baryonic Physics and When Using Polarization-only Estimators for CMB Lensing Spectra}
\label{subsubsec:baryonic}

We also consider the effect of including baryonic effects that can alter the matter power spectrum, especially at high $k$, as well as the effect of marginalizing over the shape of the kinetic Sunyaev-Zel'dovich (kSZ) power spectrum.

\begin{table}[t]
\centering

\begin{tabular}{l c c c}
\toprule
\toprule
\multicolumn{1}{l}{} & \multicolumn{2}{c}{$1\sigma$ Error} & $1\sigma$ Error Ratio \\
\cmidrule(r){2-3} \cmidrule(){4-4} 
Parameter & No Baryon Phys & Baryon Phys & with/w-out\\
\midrule
$\Omega_\mathrm{b} h^2$\dotfill & 0.000026 & 0.000026 & \quad 1.004 \\
$\Omega_\mathrm{c} h^2$\dotfill & 0.000365 & 0.000377 & \quad 1.032 \\
$H_0$\dotfill & 0.315 & 0.303 & \quad 0.962 \\
$\tau$\dotfill & 0.00447 & 0.00454 & \quad 1.014 \\
$\ln \left(10^{10} A_\mathrm{s}\right)$\dotfill & 0.00840 & 0.00855 & \quad 1.019 \\
$n_\mathrm{s}$\dotfill & 0.00181 & 0.00184 & \quad 1.015 \\
$\alpha_\mathrm{s}$\dotfill & 0.00124 & 0.00164 & \quad 1.321 \\
$N_\mathrm{eff}$\dotfill & 0.0143 & 0.0163 & \quad 1.141 \\
$\sum m_\nu$\dotfill & 0.0283 & 0.0290 & \quad 1.027 \\
$A_\mathrm{kSZ}$\dotfill & --- & 0.00140 & --- \\
$n_\mathrm{kSZ}$\dotfill & --- & 0.000864 & --- \\
$\log_{10}\left({T_\mathrm{AGN}}/{\mathrm{K}}\right)$\dotfill & --- & 0.00467 & --- \\
\bottomrule
\end{tabular}
\caption{Projected $1\sigma$ marginalized parameter error bars from a Fisher matrix analysis for CMB-HD plus DESI BAO. The first column gives the parameters; the second and third columns show the errors for a $\Lambda \mathrm{CDM} + \alpha_s + N_\mathrm{eff} + \sum m_\nu$ model and a $\Lambda \mathrm{CDM} + \alpha_s + N_\mathrm{eff} + \sum m_\nu + \log_{10}(\mathrm{T_{AGN}}/\mathrm{K}) + A_\mathrm{kSZ} + n_\mathrm{kSZ} $ model, respectively. The fourth column gives the ratio between parameter error bars when varying the additional parameters compared to keeping them fixed. A gaussian prior on $\tau$ of $\pm 0.005$ is included for both models, as well as a gaussian prior on $\log_{10}(\mathrm{T_{AGN}}/\mathrm{K})$ of $\pm 0.00468 $ in the model with bayonic physics parameters (see Section~\ref{subsubsec:baryonic} for details). $H_0$ is given in km\,s$^{-1}$\,Mpc$^{-1}$ and $\sum m_\nu$ in eV. 
For these parameter forecasts we use delensed spectra following the procedure in~\cite{MacInnis2026}.
We find that including the additional baryonic physics parameters causes the error bar on $\alpha_s$ to increase by \num{32\%}, while other parameters increase by at most \num{14\%} between models. }
\label{tab:baryonic}
\end{table}

\begin{figure}
    \centering       \includegraphics[width=.49\textwidth]{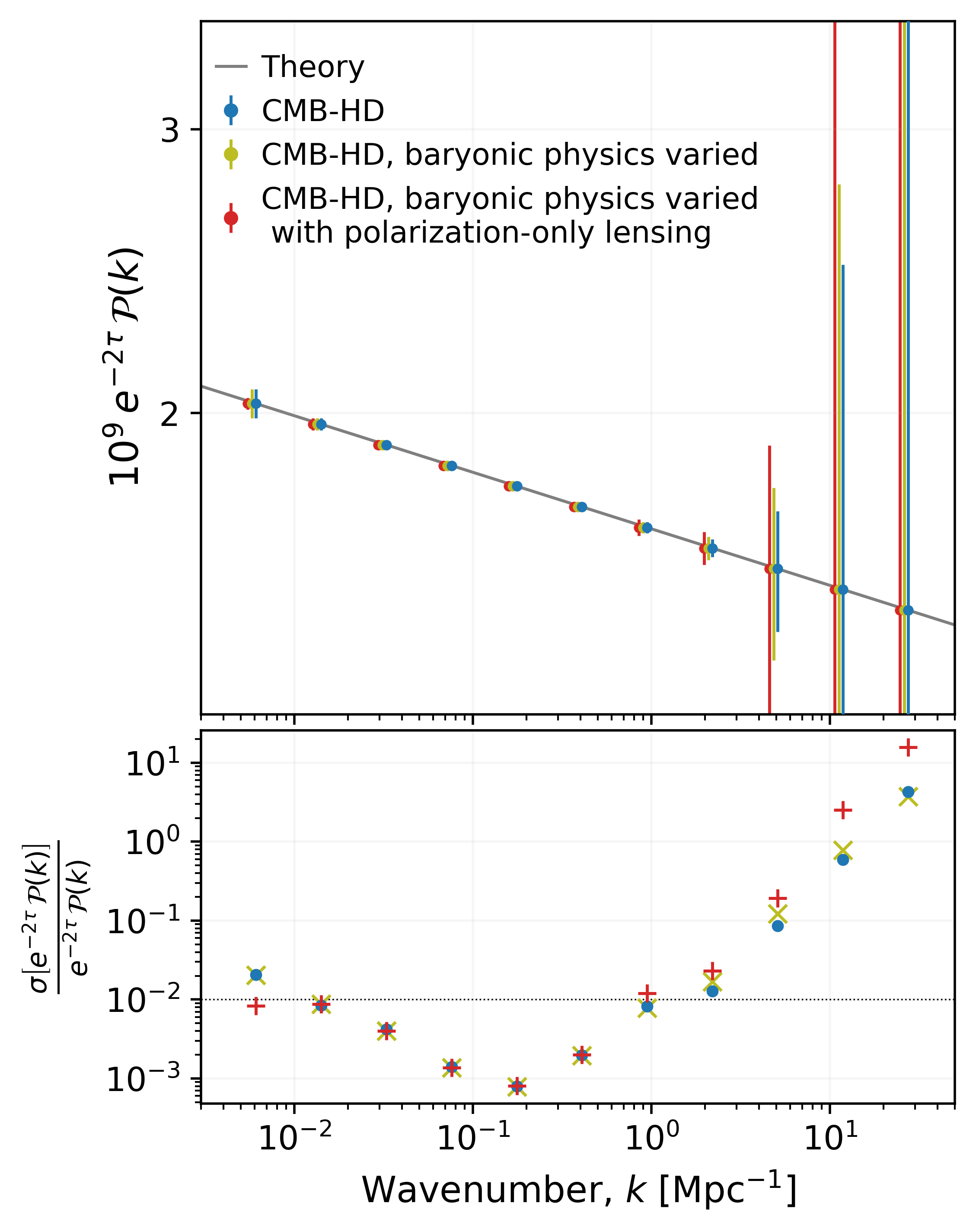}
    \caption{ Shown are projected $1\sigma$ error bars on $e^{-2\tau}\mathcal{P}(k)$ for each $k$ bin for a binned primordial power spectrum. The $e^{-2\tau}\mathcal{P}(k)$ values are centered on the fiducial primordial spectrum from Equation~\ref{eq:P(k)} evaluated with the cosmological parameters in Table~\ref{tab:fisher}.  The blue points are the same as those in Figure~\ref{fig:primordialPk} and show the Fisher forecast for CMB-HD with the three baryonic physics parameters ($\log_{10}(\mathrm{T_{AGN}}/\mathrm{K})$, $A_\mathrm{kSZ}$, and $n_\mathrm{kSZ}$) held fixed; here $\log_{10}(\mathrm{T_{AGN}}/\mathrm{K})$ is the strength of baryonic feedback, and $A_\mathrm{kSZ}$ and $n_\mathrm{kSZ}$ are the amplitude and slope of the kSZ power spectrum, respectively. The yellow points show the constraints when varying these three baryonic physics parameters in addition. We also explore switching to using CMB lensing reconstructed from only polarization estimators, which are relatively immune to potential biases from foregrounds (red points). \textit{Bottom:} The fractional uncertainty $\sigma[e^{-2\tau}\mathcal{P}(k)]/e^{-2\tau}\mathcal{P}(k)$, where the denominator is computed from Equation~\ref{eq:P(k)} with the fiducial parameter values from Table~\ref{tab:fisher}. The change from varying the additional parameters for $k < 2~\mathrm{Mpc}^{-1}$ is less than \num{5\%}, while for $2~\mathrm{Mpc}^{-1} < k < 30~\mathrm{Mpc}^{-1}$ it is less than about \num{40\%}. Switching to polarization-only estimators for the lensing reconstruction in addition increases errors by \num{45\%} for $k < 2~\mathrm{Mpc}^{-1}$ and a factor of \num{three} for $2~\mathrm{Mpc}^{-1} < k < 30~\mathrm{Mpc}^{-1}$.}
    \label{fig:HDPk-w-feedback}
\end{figure}   

\begin{figure*}[t] 
    \centering    \includegraphics[width=0.8\linewidth]{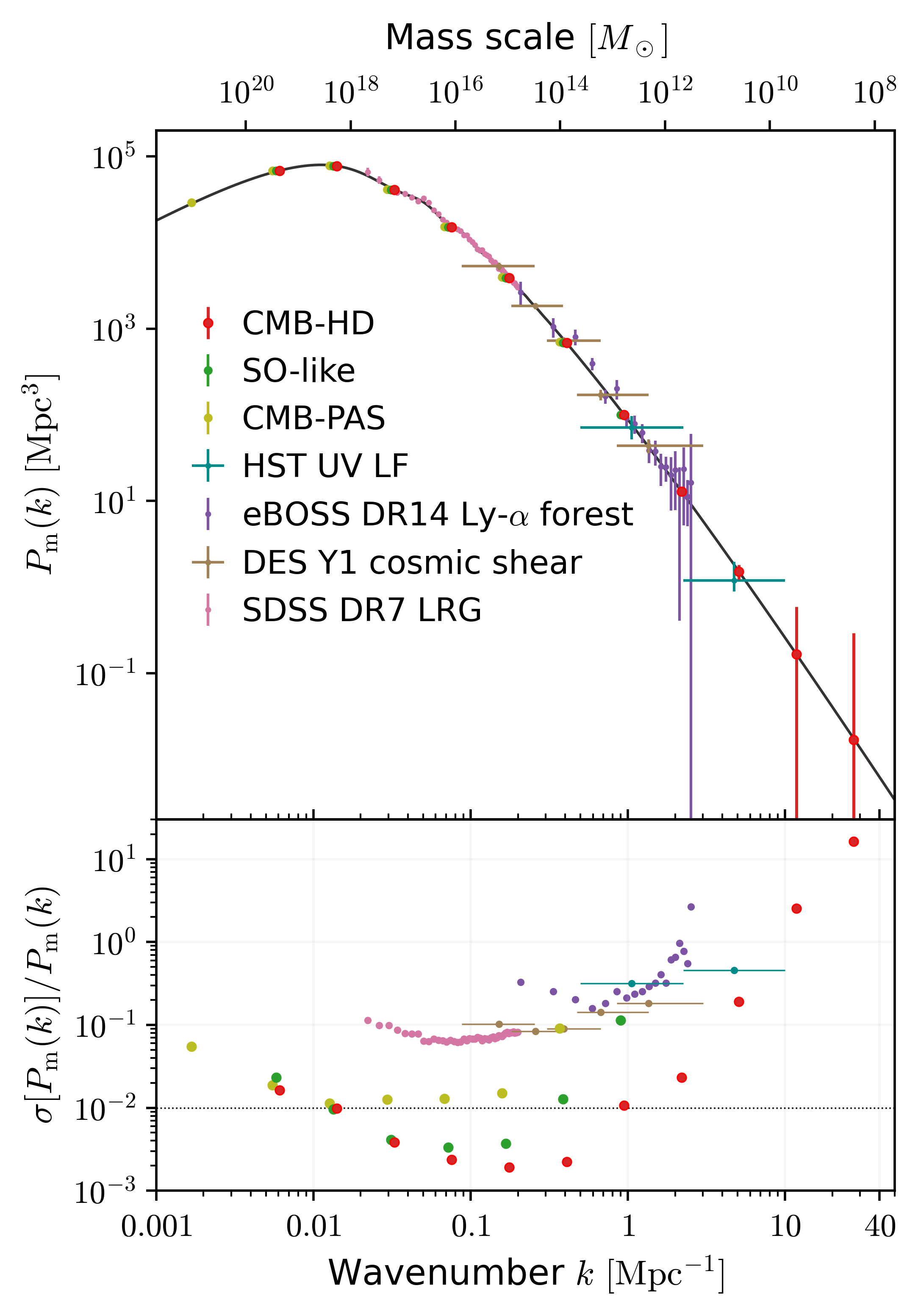}
    \caption{\textit{Top:} Shown are current constraints on the linear matter power spectrum today, $P_\mathrm{lin}(k,z=0)$, from CMB-PAS (yellow), as well as projected constraints for SO-like (green) and CMB-HD-like (red) experiments. The CMB constraints are from lensed CMB $TT, TE, EE, BB$ and CMB lensing convergence $\kappa\kappa$ spectra.  Measurements from other datasets are from a compilation by~\cite{Chabanier:2019eai}. The current and projected CMB constraints incorporate the covariance between $k$ bins and marginalize over cosmological parameters.  The CMB-HD forecasts also marginalize over baryonic physics parameters and use polarization-only estimators for the $\kappa\kappa$ spectra (see Section~\ref{subsubsec:baryonic} for details).
    \textit{Bottom:} The fractional uncertainty on the linear matter power spectrum, $\sigma(P_\mathrm{lin}(k,z=0))/P_\mathrm{lin}(k,z=0)$. We find that both SO and CMB-HD will measure $P_\mathrm{lin}(k,z=0)$ for $k < 0.2~\mathrm{Mpc}^{-1}$ to sub-percent precision, and that CMB-HD will also measure $k < 1~\mathrm{Mpc}^{-1}$ to sub-percent precision. CMB-HD will extend these measurements to $k \approx 30~\mathrm{Mpc}^{-1}$, measuring the linear matter power to within a factor of \num{ten} at these scales, within the context of a cold dark matter model. }
    \label{fig:linearPk}
\end{figure*}

To model baryonic effects, we use the \texttt{HMcode2020} model of baryonic feedback~\cite{Mead:2020vgs}, which has a single parameter, $\log_{10}(\mathrm{T_{AGN}}/\mathrm{K})$, that controls the overall strength of the feedback. For Fisher forecasts including this feedback parameter, we use the \texttt{CAMB} and \texttt{CLASS} settings given in Listings~\ref{list:camb} and~\ref{list:class}, with the change of using the \texttt{HMcode2020\_feedback} nonlinear model instead of \texttt{HMcode2016}. In \texttt{CLASS}, we do this by changing the parameter \texttt{hmcode\_version} to \texttt{`2020\_baryonic\_feedback'}; in \texttt{CAMB}, we change the parameter \texttt{halofit\_version} to \texttt{"mead2020\_feedback"}.  We then allow the parameter $\log_{10}(\mathrm{T_{AGN}}/\mathrm{K})$ to vary (called \texttt{HMCode\_logT\_AGN} in \texttt{CAMB} and \texttt{log10T\_heat\_hmcode} in \texttt{CLASS}). As was done in~\cite{MacInnis:darkmatter}, we use a fiducial value of 7.8 and a step size of 0.05 for this parameter.  As was also done in~\cite{MacInnis:darkmatter, macinnis2024}, we impose a 0.06\% gaussian prior on $\log_{10}(\mathrm{T_{AGN}}/\mathrm{K})$, as can be anticipated from CMB-HD measurements of the thermal and kinetic Sunyaev-Zel'dovich effects and CMB lensing, cross-correlated with each other; such analyses have already measured $\log_{10}(\mathrm{T_{AGN}}/\mathrm{K})$ to 6\% precision with much less constraining data~\cite{Troster:2021gsz}.

We marginalize over the shape of the kSZ power spectrum following the method described in~\cite{MacInnis:darkmatter}.
To model the kSZ effect, we use a template from~\cite{battaglia2010simulations} of the late-time kSZ effect and a model of the reionization kSZ from~\cite{Smith:2018bpn, Park:2013mv}. We use the extended template employed in~\cite{MacInnis:darkmatter}, where the templates were extended to $\ell = 20,000$ using a linear fit at the end of their existing range. We follow the approach of~\cite{Dunkley:2013vu} and normalize the kSZ power spectrum to be  $1 \mu\mathrm{K}^2$ at $\ell=3000$.  We then model the kSZ spectrum using:
\begin{equation}
        C_\ell^{\mathrm{kSZ}} = \left(\frac{\ell}{\ell_0}\right)^{n_\mathrm{kSZ}}A_{\mathrm{kSZ}}C_\ell^{kSZ,0},
\end{equation}
where $\ell_0=3000$ and $C_\ell^{kSZ,0}$ is the normalized template.  Here, $A_\mathrm{kSZ}$ is the amplitude of the kSZ power spectrum at $\ell_0$, and $n_\mathrm{kSZ}$ is its slope. We add the kSZ power spectrum to the CMB temperature power spectrum ($TT$) and vary $A_\mathrm{kSZ}$ and $n_\mathrm{kSZ}$ in our Fisher analysis using the fiducial values and step sizes given in Table~\ref{tab:fisher}.

We show in Table~\ref{tab:baryonic} the impact of varying baryonic feedback and kSZ parameters on the marginalized $1\sigma$ error bars for the cosmological parameters; we present this in the context of a 12-parameter $\Lambda \mathrm{CDM} + \alpha_s + N_\mathrm{eff} + \sum m_\nu + \log_{10}(\mathrm{T_{AGN}}/\mathrm{K}) + A_\mathrm{kSZ} + n_\mathrm{kSZ} $ model using delensed CMB spectra. We find that the error on $\alpha_s$ increases by \num{32\%} when including baryonic physics (i.e.~feedback and marginalizing over the kSZ spectrum).  However, the errors on the other cosmological parameters increase by at most \num{14\%}.

We also investigate the impact of marginalizing over these three baryonic physics parameters on the binned primordial power spectrum.  We vary the three additional parameters ($\log_{10}(\mathrm{T_{AGN}}/\mathrm{K})$, $A_\mathrm{kSZ}$, and $ n_\mathrm{kSZ}$) in addition to the four $\Lambda\mathrm{CDM}$ parameters ($\Omega_b h^2$, $\Omega_c h^2$, $\tau$, and $H_0$), and the values of $e^{-2\tau}\mathcal{P}(k)$ in the $k$ bins as given in Table~\ref{tab:P(k)}.  We show the resulting $e^{-2\tau}\mathcal{P}(k)$ constraints in Figure~\ref{fig:HDPk-w-feedback} (yellow crosses) compared to keeping the baryonic parameters fixed (blue points); the latter are the same as those shown in Figure~\ref{fig:primordialPk}. We find that for $k < 2~\mathrm{Mpc}^{-1}$ the change is less than \num{5\%}, while for $ 2~\mathrm{Mpc}^{-1} < k < 30~\mathrm{Mpc}^{-1}$ the change is less than about \num{40\%}. We also show forecasts when, in addition, we use polarization-only estimators for the CMB lensing reconstruction, which are relatively immune to extragalactic foreground biases~\cite{vanEngelen:2013rla} (red plus signs).  

\subsubsection{Constraints on the linear matter power spectrum}
\label{subsubsec:linearPk}

We translate the marginalized $1\sigma$ error bars on $e^{-2\tau}\mathcal{P}(k)$ to constraints on the binned linear matter power spectrum today, $P_\mathrm{lin}(k,z=0)$. For this, we obtain $P_\mathrm{lin}(k,z=0)$ as a derived parameter in the same way as we did for $e^{-2\tau}\mathcal{P}(k)$, following the method described in Section~\ref{sec:method-projected-Pk}. The only difference from what is described there is that we also calculate the transfer function, $\mathcal{T}(k)$, given in Eq.~\ref{eq:prim_to_lin}, for each parameter set sample that \texttt{getdist} generates. From that, we construct $P_\mathrm{lin}(k,z=0)$ for each $k$ bin using Eq.~\ref{eq:prim_to_lin}, noting that the growth function, $G(z)$, is normalized to equal one today. 

We show these constraints on the linear matter power spectrum in Figure~\ref{fig:linearPk}. We show current constraints from CMB-PAS (yellow) and projected constraints from SO-like (green) and CMB-HD-like (red) experiments.   These constraints are from lensed CMB $TT, TE, EE, BB$ and CMB lensing convergence $\kappa\kappa$ spectra. They also incorporate the covariance between $k$ bins (since all $k$ bins are varied simultaneously) and marginalize over cosmological parameters.  The CMB-HD forecasts also marginalize over baryonic physics parameters (since we vary 
$\log_{10}(\mathrm{T_{AGN}}/\mathrm{K})$, $A_\mathrm{kSZ}$, and $ n_\mathrm{kSZ}$ in addition), and they assume that polarization-only estimators are used for the $\kappa\kappa$ spectra (avoiding potential foreground biases in the lensing reconstruction).  

From the bottom panel of Figure~\ref{fig:linearPk}, we see that both SO and CMB-HD will measure $P_\mathrm{lin}(k,z=0)$ for $k < 0.2~\mathrm{Mpc}^{-1}$ to sub-percent precision. In addition, CMB-HD will measure $k < 1~\mathrm{Mpc}^{-1}$ to sub-percent precision and extend these measurements to $k \approx 30~\mathrm{Mpc}^{-1}$, measuring the linear matter power to within a factor of \num{ten} at these scales. We note that Figure~\ref{fig:linearPk} indicates how well SO and CMB-HD will measure $P_\mathrm{lin}(k,z=0)$ in the context of cold dark matter.  To constrain alternative dark matter models, it is still necessary to know how these dark matter models impact the non-linear matter power spectrum and then incorporate those theoretical spectra into the fit to the observed CMB $TT, TE, EE, BB$, and $\kappa\kappa$ spectra.

\section{Discussion and Conclusion} 
\label{sec:discussion}

In this work, we use the latest \textit{Planck}, ACT, and SPT data (CMB-PAS), plus DESI DR2 data, to constrain $n_s$ and $\alpha_s$, as well as a general binned primordial power spectrum. We find that while several of the early Universe models considered here, such as Starobinsky and Higgs inflation and the Bi-thermal Big Bang model, are ruled out at $2\sigma$ by CMB-PAS+DESI DR2 data in the context of a $\Lambda$CDM$+\alpha_s$ model, additionally freeing $N_{\rm eff}$ and $\Sigma m_\nu$ broadens the $n_s$ contours enough that these models are no longer disfavored. Freeing these parameters is well motivated, given that the sum of the neutrino masses is in tension with oscillation experiments~\cite{ACT:2025qjh}, and that $N_{\rm eff}$ need not be fixed to its standard-model value (which assumes that only neutrinos 
and no other light thermal particles have ever existed).
We further find that SO-like and, to a greater extent, CMB-HD-like experiments will have sufficient constraining power to rule out many early Universe models, including those mentioned above, even when these additional parameters are freed. Finally, all of the early Universe models we consider prefer $\alpha_s$ close to zero; if the current positive best-fit value of $\alpha_s$ holds, a CMB-HD survey would rule them out decisively.

For the binned primordial spectrum, we find consistent results between CMB-PAS and P-ACT-LB datasets, with tighter constraints from CMB-PAS for $k > 0.07\,\mathrm{Mpc}^{-1}$, owing to the inclusion of SPT data and an improved CMB lensing spectrum from a joint ACT-SPT-\textit{Planck} (APS) analysis. The CMB-PAS $e^{-2\tau}\mathcal{P}(k)$ constraints are in excellent agreement with the best-fit power-law model from \textit{Planck}~2018 data.  While CMB-PAS constrains $e^{-2\tau}\mathcal{P}(k = 0.2\,\mathrm{Mpc}^{-1})$ to \num{$0.5\%$}, we forecast that SO-like and CMB-HD-like surveys would tighten this to \num{$0.1\%$}. CMB-HD would also extend measurements of $\mathcal{P}(k)$ to $k\sim 30\,\mathrm{Mpc}^{-1}$, scales two orders of magnitude smaller than those reached by current CMB experiments, and scales not previously measured by any experiment.

CMB spectral distortions offer a complementary probe of the primordial power spectrum, as the Silk damping of acoustic modes with $50 \lesssim k \lesssim 10^4\,{\rm Mpc}^{-1}$ imprints small deviations from a blackbody spectrum~\cite{Chluba:2012we, Chluba:2012gq, Cyr:2026pic}.  Although detecting this signal is very challenging, particularly because of foreground contamination, such measurements could, in principle, complement CMB anisotropy constraints by probing even smaller scales.

This work demonstrates that current and future CMB anisotropy experiments are, and will continue to be, powerful probes of inflation and other models of the very early Universe via measurements of the primordial power spectrum.

\begin{acknowledgments}
The authors thank Renata Kallosh and Andrei Linde for helpful discussions regarding inflation models.  The authors also thank Daniel Von Thaden for testing the code presented in this work, which we make public. ZC acknowledges support from the Stony Brook URECA Summer Research Program.  EF acknowledges support under the Lourie Fellowship from the Stony Brook University Department of Physics and Astronomy. AM and NS acknowledge support from DOE award number DE-SC0025309 and the Stony Brook OVPR Seed Grant Program.  This research was supported in part by Perimeter Institute for Theoretical Physics. Research at Perimeter Institute is supported by the Government of Canada through the Department of Innovation, Science and Economic Development and by the Province of Ontario through the Ministry of Research, Innovation and Science.  The authors also note that this is not an official
Simons Observatory collaboration publication.

Additionally, the authors would like to thank Stony Brook Research Computing and Cyberinfrastructure, and the Institute for Advanced Computational Science at Stony Brook University for access to the SeaWulf computing system, made possible by grants from the National Science Foundation (award number 1531492 and Major Research Instrumentation award number 2215987), with matching funds from Empire State Development’s Division of Science, Technology, and Innovation (NYSTAR) program (contract C210148).
\end{acknowledgments}

\appendix

\section{Accuracy Settings of \texttt{CAMB} and \texttt{CLASS} and their Consistency}
\label{sec:accuracy}

In this work, we extend the compatibility of our Fisher forecast code to \texttt{CLASS} in addition to \texttt{CAMB}. To do this for CMB-HD in particular, we need to find accuracy settings for \texttt{CLASS} that are high enough so that any bias in the estimated cosmological parameter values remains below the expected statistical uncertainties. This work was performed before the release of the most recent update to \texttt{CAMB}~\cite{Lewis:2026mif}; throughout this section, we use version 1.5.9 for \texttt{CAMB} and version 3.3.0 for \texttt{CLASS}.

Accuracy settings for \texttt{CAMB} required for a CMB-HD survey were described in~\cite{macinnis2024}. We make the following small changes to what was done in~\cite{macinnis2024} and provide the updated list of accuracy settings in Listing~\ref{list:camb}.
\begin{itemize}
\item The setting \texttt{min\_l\_logl\_sampling}, which controls the multipole above which logarithmic $\ell$-sampling is used, was not included in the settings of~\cite{macinnis2024} (as that paper was written before this parameter was added to \texttt{CAMB}). Following~\cite{AtacamaCosmologyTelescope:2025nti}, we include this setting but increase its value to 10,000, which we find is needed to maintain accuracy on small scales.
\item Following~\cite{AtacamaCosmologyTelescope:2025nti}, we also explicitly set the parameter \texttt{kmax}.  However, we set it to 100, which is higher than the value of 10 used in~\cite{AtacamaCosmologyTelescope:2025nti}, as CMB-HD will probe scales beyond $k = 10$~Mpc$^{-1}$.  (In ~\cite{MacInnis:darkmatter}, this parameter was set to 1000, but we find this is higher than needed.)
\item In previous work, we let \texttt{CAMB} optimize \texttt{k\_per\_logint}.  However, in this work, we set \texttt{k\_per\_logint = 130} following~\cite{AtacamaCosmologyTelescope:2025nti}.
\item As discussed in Section~\ref{sec:method-current}, we set \texttt{num\_massive\_neutrinos = 3} for three massive degenerate neutrino species, in contrast to \texttt{num\_massive\_neutrinos = 1} used in~\cite{macinnis2024,MacInnis:darkmatter}.
\item For clarity, we explicitly set \texttt{bbn\_predictor} to its default value, which uses PRIMAT 2021.  This file (and its \texttt{CLASS} equivalent) contains the tabulated primordial abundances predicted by the Big Bang Nucleosynthesis code PRIMAT~\cite{Pitrou:primat} of He-4 and Deuterium as a function of $\Omega_b h^2$ and the number of extra relativistic degrees of freedom  relative to the standard value, $\Delta N_\mathrm{eff}$. \texttt{CAMB} also provides BBN predictions from different versions of PRIMAT and from  PArthENoPE~\cite{Gariazzo:parthenope}.
\end{itemize}

\begin{figure*}[t]
\begin{lstlisting}[caption={\texttt{CAMB} Accuracy Settings Needed for CMB-HD},label={list:camb}, captionpos=t, language=Python]
import camb
import numpy as np
lmax = 24000 
pars = camb.CAMBparams()
pars.set_cosmology(H0=67.36, ombh2=0.02237, omch2=0.1200, tau=0.0544, num_massive_neutrinos=3, mnu=0.06, nnu=3.044, bbn_predictor="PRIMAT_Yp_DH_ErrorMC_2021.dat")
pars.set_classes(recombination_model = "Recfast")
pars.InitPower.set_params(As=np.exp(3.044) * 1e-10, ns=0.9649)
pars.set_matter_power(kmax=100, k_per_logint=130)
pars.set_for_lmax(lmax+500, lens_potential_accuracy=30, lens_margin=2050)
pars.set_accuracy(AccuracyBoost=1.1, lSampleBoost=3.0, lAccuracyBoost=3.0, DoLateRadTruncation=False, min_l_logl_sampling=10000)
pars.NonLinear = camb.model.NonLinear_both
pars.NonLinearModel.set_params(halofit_version="mead2016")
\end{lstlisting}
\end{figure*}

To determine the \texttt{CLASS} accuracy settings, we take the settings given in~\cite{AtacamaCosmologyTelescope:2025nti} as a starting point and make the following changes:
\begin{itemize}
\item We increase \texttt{P\_k\_max\_h/Mpc} to 500 from 100. This setting controls the maximum wavenumber $k$ used for the matter power spectrum.  We show in Figure~\ref{fig:bias} that this setting is needed to keep parameter biases below statistical errors. 
\item While we use the \texttt{CosmoRec}~\cite{chluba:cosmorec_2, chluba:cosmorec_1} recombination model and \texttt{CAMB} for all our constraints from current data, \texttt{CLASS} currently has no \texttt{CosmoRec} option. Thus, for our projected constraints, we use a recalibrated \texttt{RECFAST} recombination model in both \texttt{CAMB} and \texttt{CLASS}, since it is available in the packaged versions of both codes, which allows us to check their consistency.  Note that the \texttt{HyRec} recombination model is the default option for \texttt{CLASS}.  We find that for parameter {\it{error bars}}, for the CMB-HD accuracy settings, there is less than a 0.7\% change between using \texttt{CosmoRec} and \texttt{RECFAST} for \texttt{CAMB}, and less than a 2\% change between using \texttt{HyRec} and \texttt{RECFAST} for \texttt{CLASS}.  We expect larger differences for parameter {\it{biases}}, and thus note that it is important to have high accuracy recombination modeling. 
\item  We set \texttt{N\_ur} as required for $N_\mathrm{eff}=3.044$ and three massive neutrinos, computed from the equation \texttt{N\_ur}~=~$N_\mathrm{eff} - 1.0131966(n_\nu)$, where $n_\nu = 3$, as discussed in Section~\ref{sec:method-projected}. We specify the neutrino masses (with $\sum m_\nu$~=~0.06~eV) by passing \texttt{m\_ncdm = `0.02, 0.02, 0.02'}. We also set \texttt{N\_ncdm = 3} and \texttt{deg\_ncdm = `1, 1, 1'}. These settings ensure that three massive neutrinos are used.
\item We set \texttt{N\_ur} explicitly, as opposed to setting \texttt{N\_eff}. Calculating \texttt{N\_ur} from our value of $N_\mathrm{eff}$ using the equation above and then setting \texttt{N\_ur} allows \texttt{CAMB} and \texttt{CLASS} spectra to agree to within 0.2\% below $\ell = 15,000$ (as shown in Figure~\ref{fig:camb_over_class}),while setting \texttt{N\_eff} explicitly gives agreement to within 0.5\%. 
\item We set \texttt{delta\_l\_max} to 2050 for parity with our \texttt{CAMB} settings, where the comparable setting is \texttt{lens\_margin = 2050} (or \texttt{lens\_output\_margin = 2050} for more recent versions of \texttt{CAMB}). These settings control the maximum multipole used when calculating lensed power spectra by specifying how much to increase it beyond the given maximum multipole.
\item We set the \texttt{sBBN file} to PRIMAT 2021 for consistency with \texttt{CAMB}. This is the same PRIMAT bbn file as used for \texttt{CAMB}, but modified to be in the format required for \texttt{CLASS}. Notably, only the He-4 abundance is kept for \texttt{CLASS}; the deuterium abundance is discarded. While \texttt{CLASS} contains files with BBN data from \texttt{PRIMAT}, these files are from 2017 and 2025.  To ensure maximum consistency between the codes, we follow~\cite{AtacamaCosmologyTelescope:2025nti} and use the 2021 \texttt{PRIMAT} BBN file in \texttt{CAMB}, modified to be consistent with the formatting expected by \texttt{CLASS}.
\end{itemize}

We also make the following changes to the standard \texttt{CLASS} implementation:

\begin{itemize}
    \item In the \texttt{CLASS} code, in the \texttt{source/lensing.c} file,\footnote{\url{https://github.com/lesgourg/class_public/blob/v3.3.4/source/lensing.c\#L124}} the variable \texttt{num\_mu} and \texttt{index\_mu} have to be changed to \texttt{long long} type.  In addition, \texttt{icount} must be an \texttt{unsigned long long} for the code to work at the multipoles we consider. Otherwise, the program encounters an overflow error and attempts to allocate memory with a negative size. This change is displayed in Listing~\ref{list:class_variable_change}.
    
    \item By default, \texttt{CLASS} does not allow \texttt{N\_ur} to go negative. This means that when \texttt{N\_ncdm} or  \texttt{deg\_ncdm} is set to three so that one has three massive neutrinos, \texttt{CLASS} will not allow $N_\mathrm{eff}$ to be varied below approximately 3.039. This is an issue when attempting to do an MCMC run while varying $N_\mathrm{eff}$, since it prevents the corresponding portion of parameter space from being explored. This is also an issue when using a 5\% step size for Fisher derivative calculations. To circumvent this issue, the line that checks for negative values of \texttt{N\_ur} can be commented out. This is line 2470 in the \texttt{source/input.c} file for \texttt{CLASS} version 3.3.4.\footnote{\url{https://github.com/lesgourg/class_public/blob/v3.3.4/source/input.c\#L2470}} This change is displayed in Listing~\ref{list:class_commented_out}. This allows the full parameter space to be explored by an MCMC run and lets Fisher estimates use the step size listed in Table~\ref{tab:fisher} for $N_\mathrm{eff}$. We verify the accuracy of \texttt{CLASS} spectra produced by values of $N_\mathrm{eff}$ that would have produced this error by comparing them with \texttt{CAMB}, as \texttt{CAMB} does not impose this limitation on $N_\mathrm{eff}$. Figure~\ref{fig:camb_class_triangle} and Table \ref{tab:camb_class_errs} show that both Fisher matrices and MCMC runs computed using \texttt{CAMB} and \texttt{CLASS}, after this \texttt{CLASS} change, agree well with each other.
\end{itemize}

\begin{figure}[t]
\begin{lstlisting}[caption={The \texttt{CLASS} changes we made with the original line commented out. This code is located in the \texttt{source/lensing.c} file. In \texttt{CLASS} version 3.3.4, the unedited code is at line 124.},label={list:class_variable_change}, captionpos=t, language=C, firstnumber=124]
unsigned long long icount; /* was int num_mu,index_mu,icount */
long long num_mu, index_mu;
int l;
double ll;
\end{lstlisting}
\end{figure}

\begin{figure}[t]
\begin{lstlisting}[caption={This line of code in \texttt{CLASS} gives an error when \texttt{N\_ur} is negative; thus, we commented it out, which we find works fine. This code is located in the \texttt{source/input.c} file; in \texttt{CLASS} version 3.3.4, it is line 2469.},label={list:class_commented_out}, captionpos=t, language=C, firstnumber=2470]
/* we comment this line out: 
class_test(pba->Omega0_ur<0, errmsg,"You cannot set the density of ultra-relativistic relics (dark radiation/neutrinos) to negative values. You might have input a total Neff smaller than what your massive neutrinos require minimally (around 1.02 * N_ncdm * deg_ncdm).");
*/    
\end{lstlisting}
\end{figure}

We list the final accuracy settings for \texttt{CLASS} in Listing~\ref{list:class}.  In Figure~\ref{fig:camb_over_class}, we show the agreement between \texttt{CAMB} and \texttt{CLASS} power spectra generated using the settings given in Listings~\ref{list:camb} and~\ref{list:class}, respectively.  We show the ratio of \texttt{CAMB} lensed power spectra to \texttt{CLASS} for $TT$ (blue), $EE$ (orange), $BB$ (green), and $\kappa \kappa$ (red); $TE$ is not displayed since it crosses zero at many points.  We find that all \texttt{CAMB} and \texttt{CLASS} power spectra agree to within 0.5\% out to $\ell = 20,000$ using these accuracy settings. \\

To estimate the bias on cosmological parameters, we follow the method in~\cite{macinnis2024}.  In particular, the bias on parameter $\theta_\alpha$ is given by
\begin{equation}
        \Delta \theta_\alpha = \sum_\beta F^{-1}_{\alpha\beta} \sum_{\ell_b, \ell_b'} \frac{\partial C_{\ell_b}^\mathrm{fid}}{\partial \theta_\beta} \mathbb{C}^{-1}_{\ell_b \ell_b'} \left(C_{\ell_b'}^\mathrm{true} - C_{\ell_b'}^\mathrm{fid}\right).
\label{eq:bias}
\end{equation}
as described in~\cite{Huterer2004,Amara2007,Bernal2020}. Here, $\alpha$ and $\beta$ are indices corresponding to the parameters; $C_{l_b}^{\mathrm{true}}$ are the ``true Universe'' spectra at a multipole $\ell_b$; $C_{l_b}^{\mathrm{fid}}$ are the spectra at a multipole $\ell_b$ computed with some fiducial accuracy settings; $F^{-1}_{\alpha \beta}$ is the $\alpha, \beta$ element of the inverse Fisher matrix calculated with the fiducial settings; and $\mathbb{C}^{-1}_{\ell_b \ell_b'}$ is the $\ell_b, \ell_{b'}$ element of the data covariance matrix. 

For the CMB power spectra of the ``true Universe,'' we choose to use \texttt{CAMB} spectra with the accuracy settings described in Listing~\ref{list:camb}, along with the additional changes of \texttt{AccuracyBoost = 3.0}, \texttt{lAccuracyBoost = 5.0}, \texttt{lens\_potential\_accuracy = 40.0}, and \texttt{lSampleBoost = 5.0}. These settings are much more computationally intensive than our baseline settings, but they are also more accurate.  We aim for \texttt{CAMB} and \texttt{CLASS} accuracy settings that result in parameter biases of less than $0.5\sigma$ (less than half the statistical error bar).

In Figure~\ref{fig:bias}, we show the bias on each parameter from Eq.~\ref{eq:bias}, expressed in units of the parameter uncertainty in a $\Lambda\mathrm{CDM} + N_\mathrm{eff} + \sum m_\nu$ model.  The left panel shows parameter biases using \texttt{CAMB} as a function of the \texttt{lens\_potential\_accuracy} setting, and the right
panel shows those from \texttt{CLASS} as a function of the \texttt{P\_k\_max\_h/Mpc} setting; these settings were chosen because they produced large changes in the spectra when altered while keeping the other settings fixed.  For this Figure, we only vary these settings and hold the others fixed to those in Listings~\ref{list:camb} and~\ref{list:class}.  We see that for \texttt{lens\_potential\_accuracy = 30} and \texttt{P\_k\_max\_h/Mpc = 500}, the bias for all cosmological parameters is less than $0.5\sigma$.  For the higher accuracy settings shown, there is minimal additional improvement.  Thus, we adopt these as our baseline settings.

\begin{figure}[t]
    \centering
    \includegraphics[width=\columnwidth]{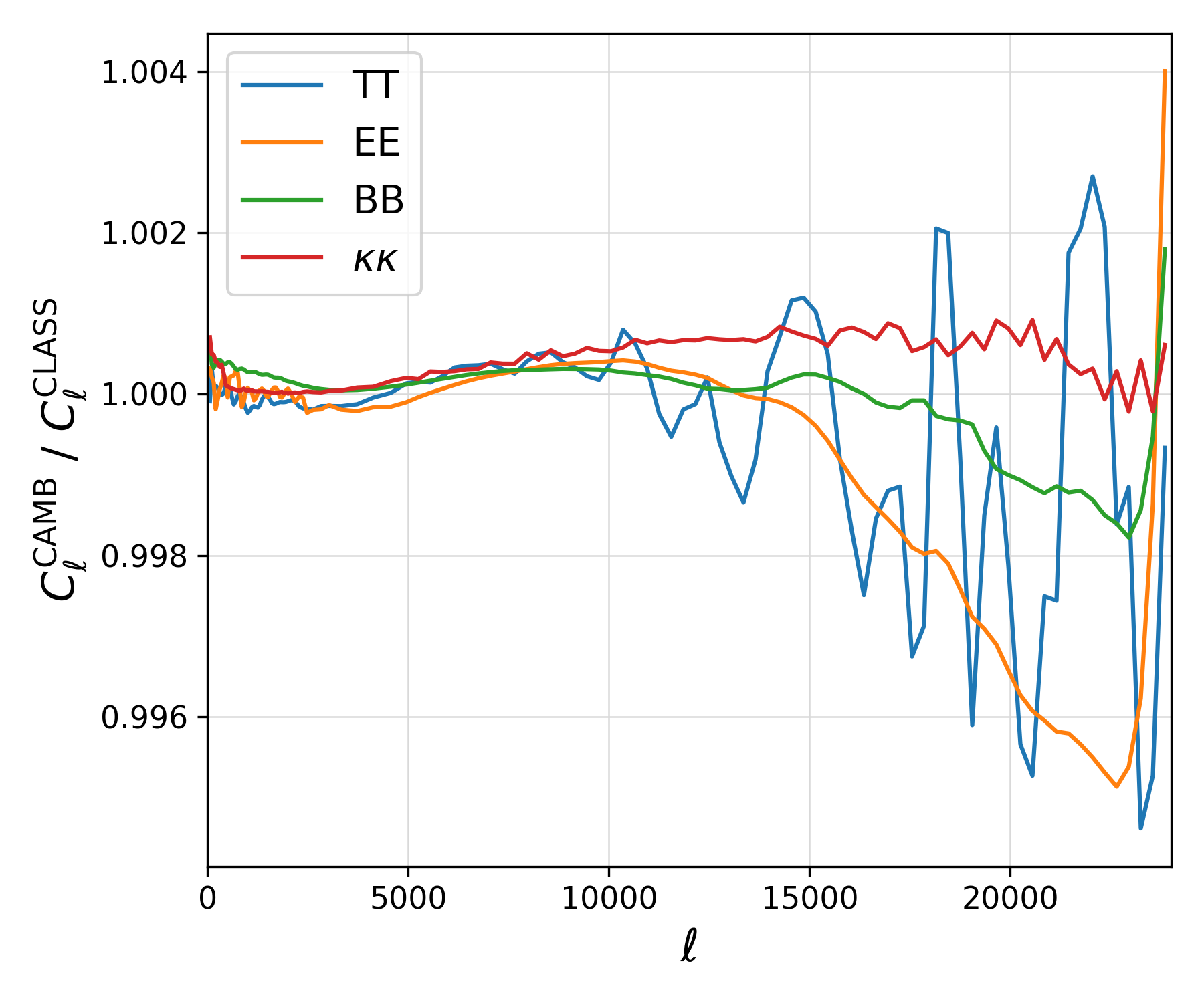}
    \caption{Ratio of lensed CMB $TT$ (blue), $EE$ (orange), $BB$ (green) and CMB lensing convergence $\kappa\kappa$ (red) binned power 
    spectra computed with \texttt{CAMB} to those computed with \texttt{CLASS}. The $TE$ spectra is not shown since it crosses zero in multiple places. The power spectra were calculated using the fiducial cosmology listed in Table~\ref{tab:fisher} and the accuracy settings provided in Listings~\ref{list:camb} and~\ref{list:class} for \texttt{CAMB} and \texttt{CLASS}, respectively. The two sets of power spectra agree to within $0.5\%$ out to $\ell = 20{,}100$, which is the maximum multipole considered for all calculations in this work.}
    \label{fig:camb_over_class}
\end{figure}

\begin{figure*}[t]
\begin{lstlisting}[caption={\texttt{CLASS} Accuracy Settings Needed for CMB-HD},label={list:class}, captionpos=t, language=Python]
from classy import Class
lmax = 24000
settings = {'H0' : 67.36, 
'omega_b' :0.02237 , 
'omega_cdm' :0.1200, 
'tau_reio' :0.0544,
'n_s': 0.9649, 
'ln_A_s_1e10': 3.044,
'N_ncdm': 3,
'm_ncdm': '0.02,0.02,0.02',
'deg_ncdm': '1,1,1'
'N_ur': 0.004410200000000142,
'T_cmb': 2.7255,
'YHe': 'BBN',
'sBBN file': "PRIMAT21_class_format.dat",
'non_linear': 'hmcode',
'hmcode_version': '2016',
'recombination': 'recfast',
'lensing': 'yes',
'output': 'lCl,tCl,pCl,mPk',
'modes': 's',
'l_max_scalars': lmax + 500,
'delta_l_max': 2050,
'P_k_max_h/Mpc': 500,
'l_logstep': 1.025,
'l_linstep': 20,
'perturbations_sampling_stepsize': 0.05,
'l_switch_limber': 30.,
'hyper_sampling_flat': 32.,
'l_max_g': 40,
'l_max_ur': 35,
'l_max_pol_g': 60,
'ur_fluid_approximation': 2,
'ur_fluid_trigger_tau_over_tau_k': 130.,
'radiation_streaming_approximation': 2,
'radiation_streaming_trigger_tau_over_tau_k': 240.,
'hyper_flat_approximation_nu': 7000.,
'transfer_neglect_delta_k_S_t0': 0.17,
'transfer_neglect_delta_k_S_t1': 0.05,
'transfer_neglect_delta_k_S_t2': 0.17,
'transfer_neglect_delta_k_S_e': 0.17,
'accurate_lensing': 1,
'start_small_k_at_tau_c_over_tau_h': 0.0004,
'start_large_k_at_tau_h_over_tau_k': 0.05,
'tight_coupling_trigger_tau_c_over_tau_h': 0.005,
'tight_coupling_trigger_tau_c_over_tau_k': 0.008,
'start_sources_at_tau_c_over_tau_h': 0.006,
'l_max_ncdm': 30,
'tol_ncdm_synchronous': 1.e-6,}
M = Class()
M.empty()
M.set(settings)
\end{lstlisting}
\end{figure*}

\begin{figure*}[t]
    \centering
    \includegraphics[width=\linewidth]{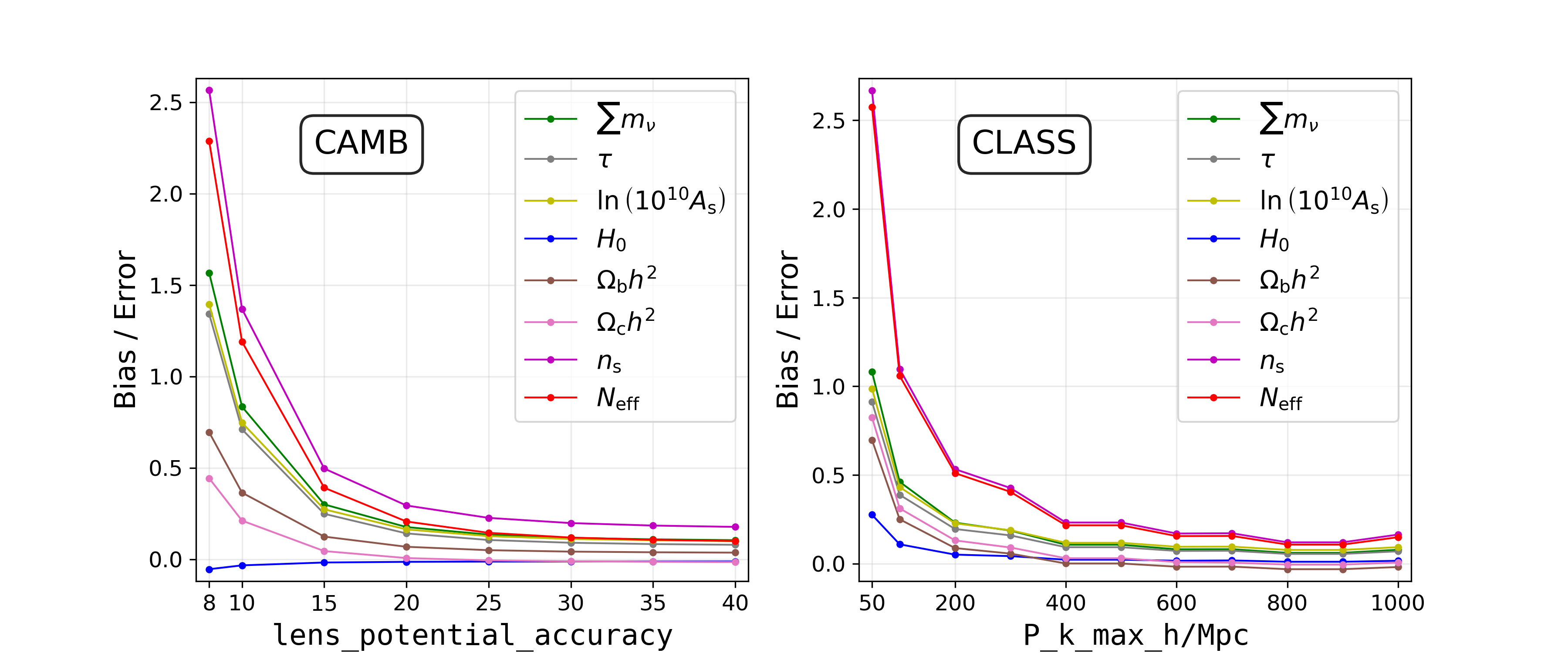}
    \caption{The bias on the cosmological parameters shown as a fraction of their projected $1\sigma$ uncertainties for CMB-HD in the context of a $\Lambda\mathrm{CDM}+N_\mathrm{{eff}}+{m_\nu}$ model.  This is shown for \texttt{CAMB} (left panel) and \texttt{CLASS} (right panel) using the accuracy settings given in Listings~\ref{list:camb} and \ref{list:class}, respectively, with the exception of the setting on the x-axis, which is allowed to vary. The \texttt{lens\_potential\_accuracy} setting for \texttt{CAMB} and the \texttt{P\_k\_max\_h/Mpc} setting for \texttt{CLASS} were chosen as they give significant changes to the spectra and biases when varied. The parameter biases are calculated using Equation~\ref{eq:bias}.  In both panels, we define the  ``true Universe'' as \texttt{CAMB} spectra using the settings in Listing~\ref{list:camb} but increasing the accuracy further by setting \texttt{AccuracyBoost = 3.0}, \texttt{lAccuracyBoost = 5.0},  \texttt{lens\_potential\_accuracy = 40.0}, and \texttt{lSampleBoost = 5.0}.  We find that \texttt{lens\_potential\_accuracy = 30} and \texttt{P\_k\_max\_h/Mpc = 500} yield biases below $0.5\sigma$ for all parameters, with minimal improvement for higher settings.
    Thus, we choose these settings as our baseline throughout this work.
    }
    \label{fig:bias}
\end{figure*}

\begin{comment}
 \begin{table}[t]
    \begin{center}
    \begin{tabular}{c@{\hskip 2em} c@{\hskip 1em} c}
      \toprule
      \toprule
       %Calculation &  \multicolumn{2}{c}{Time }    \\
       %\cmidrule(lr){2-3}
      \multicolumn{3}{c}{Computation Time }    \\
       \cmidrule(lr){1-3}
        & CAMB  &  CLASS  \\
      \bottomrule%\midrule
      Single call for power spectra  & 7 seconds & 51 seconds  \\
      \midrule
      MCMC (6 parameters)& 51 hours & 99 hours \\
      
      MCMC (9 parameters)  & 248 hours & 281 hours \\
      \midrule
      \bottomrule
    \end{tabular}    
    \caption{Computation times for a single compute call of \texttt{CAMB} and \texttt{CLASS} power spectra using the required accuracy settings for CMB-HD listed in Listings \ref{list:camb} and \ref{list:class}, respectively.  In addition, we give the time it took for MCMC runs to reach to R-1 = 0.01 with both codes for the six-parameter $\Lambda$CDM and the nine-parameter $\Lambda$CDM+$N_\mathrm{eff}$+$\sum m_\nu$ models. Computations were performed on the SEAWULF hbm-long-96core compute nodes, using 6 chains for each MCMC run. Each chain received its own node with 96 CPU cores and 384 GB of memory. We find \texttt{CLASS} computation times are significantly longer than those for \texttt{CAMB}.  For more complex models, emulators will likely be required to speed up MCMC runs. Note times can be faster on other machines. \NS{Update with latest numbers.}} 
    \label{tab:times}
    \end{center}
\end{table}
\end{comment}   

\section{Stability of Forecasts: \texttt{CAMB} vs. \texttt{CLASS} and Fisher vs. MCMC}
\label{sec:stabillity}

In Table~\ref{tab:camb_class_errs} and Figure~\ref{fig:camb_class_triangle}, we show the agreement between parameter uncertainties from Fisher estimates and MCMC runs, as well as from \texttt{CAMB} and \texttt{CLASS}.  Table~\ref{tab:camb_class_errs} indicates that \texttt{CAMB} and \texttt{CLASS} Fisher estimates of parameter errors match to within \num{5\%}, while Fisher and MCMC errors match to within \num{7\%} for \texttt{CAMB} and \num{4\%} for \texttt{CLASS}. For the MCMC runs, we supplied a proposal matrix to \texttt{Cobaya}, which shortened convergence times.  We adopted two methods to obtain proposal matrices: 1)~run a lower accuracy MCMC using the inverted Fisher matrix as a proposal matrix, and 2)~run a short MCMC with the full accuracy settings, stopping it at $R-1 = 0.3$; then use the covariance matrix from this shorter run as an input proposal matrix for a longer run.

\begin{table*}[t]
\centering
\begin{tabular}{l@{\hskip 1.5em} c@{\hskip 2em} c c c c c@{\hskip 2em} c c@{\hskip 1.5em} c@{\hskip 1em} c}
\toprule
\toprule
\multicolumn{1}{l}{} & \multicolumn{2}{c}{Fisher $1\sigma$ Error} & & \multicolumn{1}{c}{Fisher $1\sigma$ Error Ratio} & &\multicolumn{2}{c}{MCMC $1\sigma$ Error} &  & \multicolumn{2}{c}{Fisher/MCMC}  
\\
\cmidrule(lr){2-3} \cmidrule(){5-5} \cmidrule(lr){7-8} \cmidrule(){10-11} 
Parameter & CAMB  & CLASS & & CAMB / CLASS & & CAMB & CLASS&&  CAMB & CLASS\\ 
\midrule
$\Omega_\mathrm{b} h^2$\dotfill & 0.000026 & 0.000026 &  & 0.999 &  & 0.000026 & 0.000026 &  & 1.005 & 1.012 \\
$\Omega_\mathrm{c} h^2$\dotfill & 0.000388 & 0.000392 &  & 0.992 &  & 0.000385 & 0.000408 &  & 1.007 & 0.960 \\
$H_0$\dotfill & 0.317 & 0.311 &  & 1.020 &  & 0.299 & 0.318 &  & 1.061 & 0.977 \\
$\tau$\dotfill & 0.00454 & 0.00456 &  & 0.995 &  & 0.00435 & 0.00452 &  & 1.044 & 1.009 \\
$\ln \left(10^{10} A_\mathrm{s}\right)$\dotfill & 0.00847 & 0.00852 &  & 0.994 &  & 0.00806 & 0.00831 &  & 1.051 & 1.025 \\
$n_\mathrm{s}$\dotfill & 0.00193 & 0.00193 &  & 0.999 &  & 0.00194 & 0.00190 &  & 0.996 & 1.016 \\
$\alpha_\mathrm{s}$\dotfill & 0.00129 & 0.00123 &  & 1.046 &  & 0.00127 & 0.00122 &  & 1.012 & 1.006 \\
$N_\mathrm{eff}$\dotfill & 0.0153 & 0.0153 &  & 0.998 &  & 0.0151 & 0.0148 &  & 1.012 & 1.032 \\
$\sum m_\nu$\dotfill & 0.0283 & 0.0281 &  & 1.005 &  & 0.0265 & 0.0287 &  & 1.067 & 0.981 \\
\bottomrule
\end{tabular}
\caption{Comparison of the projected marginalized $1\sigma$ parameter errors for a $\Lambda\mathrm{CDM}+\alpha_s+N_\mathrm{eff}+\sum m_\nu$ model from mock CMB-HD plus DESI BAO data. The first column lists the parameters. The second and third columns present two sets of Fisher forecasts: one using \texttt{CAMB}, and one using \texttt{CLASS}, using the accuracy settings described in Appendix~\ref{sec:accuracy}. A Gaussian prior of $\sigma(\tau) = 0.005$ is applied in both cases (see Table~\ref{tab:fisher}). The fourth column gives the ratio of \texttt{CAMB} to \texttt{CLASS} Fisher errors; we find agreement to within \num{5\%} between the two codes. Columns five and six display the $1\sigma$ marginalized errors from \texttt{CAMB} and \texttt{CLASS} MCMC runs using \texttt{Cobaya} for CMB-HD plus DESI BAO mock data. Column seven displays the ratio of Fisher errors to MCMC errors for \texttt{CAMB}, with agreement to within \num{7\%}.  Column eight shows the ratio of Fisher errors to MCMC errors for \texttt{CLASS}, with agreement to within \num{4\%}. }
\label{tab:camb_class_errs}
\end{table*}

\begin{figure*}[t]
    \centering
    \includegraphics[width=\linewidth]{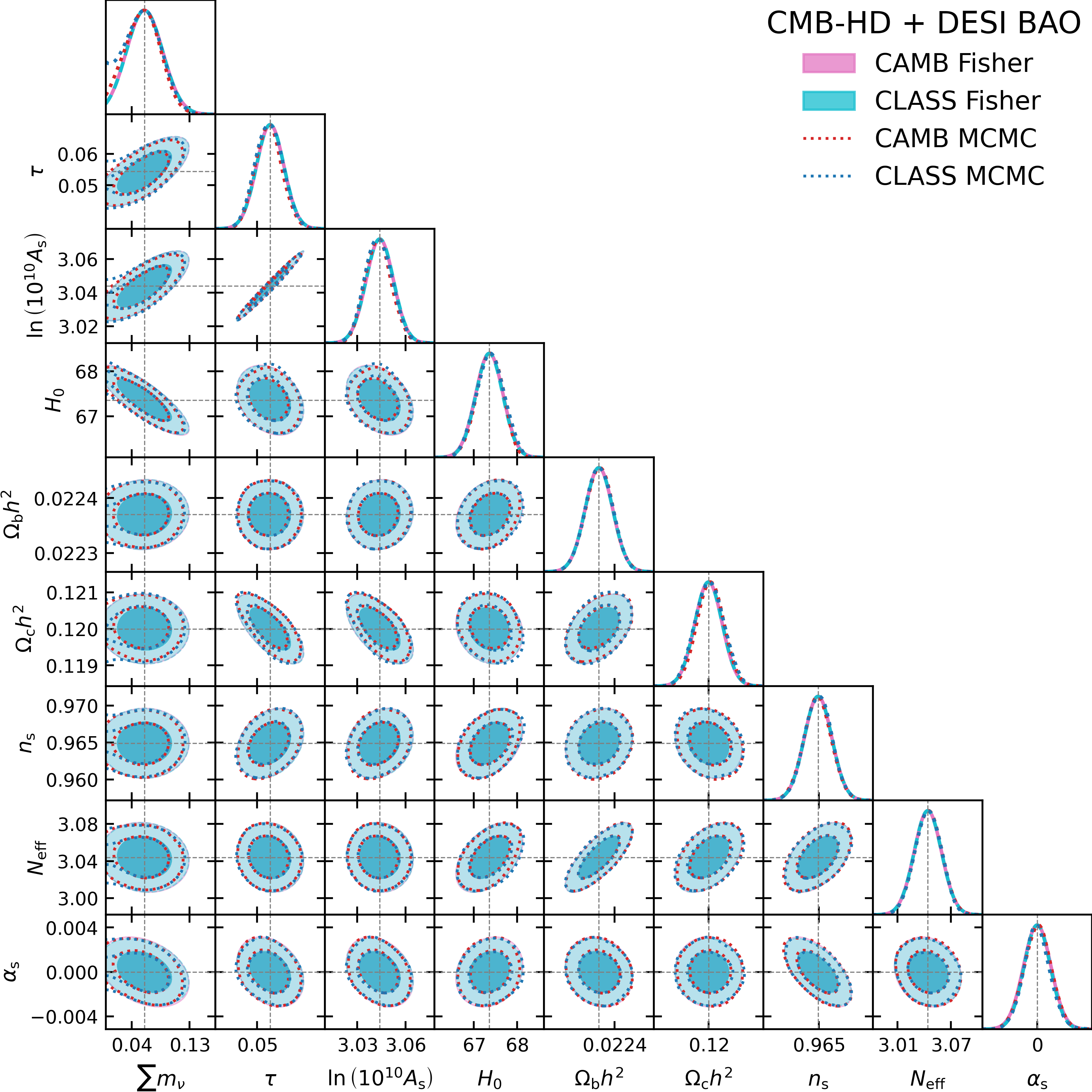}
    \caption{Comparison of parameter forecasts from mock CMB-HD plus DESI BAO data using the methods described in Section~\ref{sec:method-projected} for the nine-parameter $\Lambda\mathrm{CDM}+\alpha_s+N_\mathrm{eff}+\sum m_\nu$ model. We show the results from Fisher forecasts as the solid contours, and the results from MCMC chains as the dotted lines; for both forecasting methods, we show the constraints when using either \texttt{CAMB} (pink solid and red dotted) or \texttt{CLASS} (cyan solid and blue dotted). The gray dashed lines indicate the fiducial parameter values from Table~\ref{tab:fisher}. All MCMC chains shown have Gelman-Rubin statistic $R-1 < 0.01$  after removal of a 50\% burn-in fraction. We find that all methods produce consistent parameter constraints, demonstrating the agreement between the Fisher and MCMC methods and between \texttt{CAMB} and \texttt{CLASS}. We quantify this agreement further in Table~\ref{tab:camb_class_errs}. }
    \label{fig:camb_class_triangle}
\end{figure*}

\section{Primordial Power Spectrum with Finer Binning and Addition of DESI BAO}
\label{sec:AppendixPk}

Since constraints on the binned primordial power spectrum are sensitive to the binning scheme used, we show in Figure~\ref{fig:30bin} the constraints from the P-ACT-LB and CMB-PAS datasets using the finer 30-bin scheme adopted in~\cite{AtacamaCosmologyTelescope:2025nti}.  We reproduce the results of the official P-ACT-LB chains (shown in black) and still find that CMB-PAS improves over P-ACT-LB, as also shown in Figure~\ref{fig:primordialPk} for the seven-bin case. For these results, we use the same priors as were used in~\cite{AtacamaCosmologyTelescope:2025nti}, given in Table 5 of Appendix C.  We simultaneously vary the values in the 30 $k$ bins plus the four additional $\Lambda\mathrm{CDM}$ parameters: $\Omega_bh^2$, $\Omega_ch^2$, $\tau$, and $\theta_\mathrm{MC}$.

We also investigate the impact of including DESI BAO data and find no significant improvement in either the 30-bin (Figure~\ref{fig:30bin}) or seven-bin (Figure~\ref{fig:comparePk7bin}) cases.  In Table~\ref{tab:pk-errors}, we provide the marginalized mean values and $1\sigma$ errors on $e^{-2\tau}\mathcal{P}(k)$ for a binned primordial power spectrum from the P-ACT-LB (second column), CMB-PAS (third column), and CMB-PAS+DESI DR2 (fourth column) datasets.  We see consistent results between P-ACT-LB and CMB-PAS, with slightly smaller errors for the CMB-PAS dataset, which includes SPT data~\cite{SPT-3G:2025bzu} and an improved CMB lensing spectrum from a joint ACT-SPT-{\it{Planck}} (APS) analysis~\cite{ACT:2025qjh}.  We see that adding DESI DR2 to CMB-PAS has minimal impact. This is consistent with the results of~\cite{Nerval:2026iev}, which observed small changes to the P-ACT-LB constraints on a primordial power spectrum when including DESI DR2 data.  We also show forecasts for SO-like and CMB-HD-like surveys and note that CMB-HD constrains $k$ values almost two orders of magnitude higher than current surveys.

\begin{figure*}[t]
    \centering
    \begin{minipage}[t]{.49\textwidth}       \includegraphics[width=\textwidth]{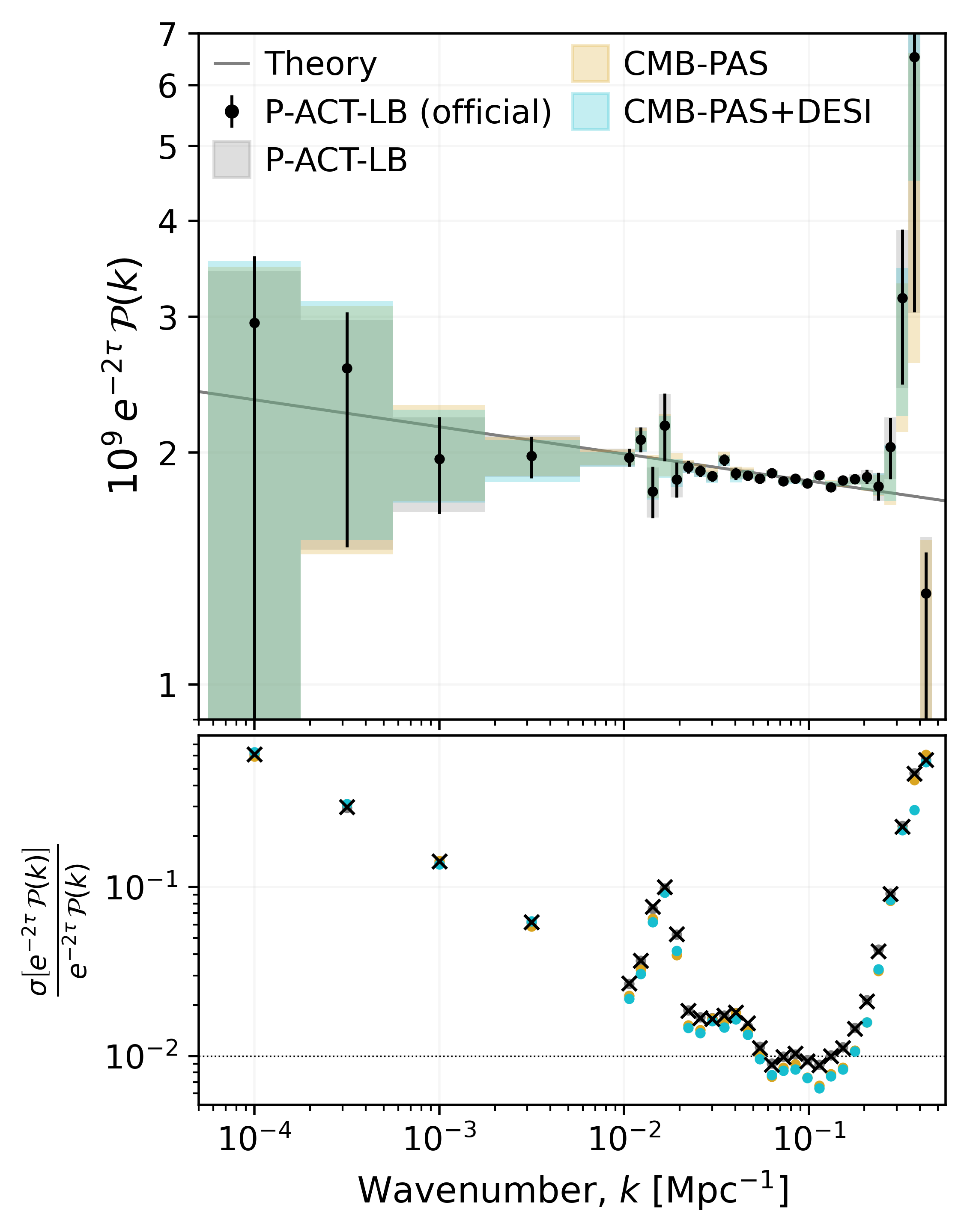}
    \caption{Constraints on a binned primordial power spectrum using the finer 30-bin scheme of~\cite{AtacamaCosmologyTelescope:2025nti}. \textit{Top:} Shown are marginalized means and $1\sigma$ error bars on $e^{-2\tau}\mathcal{P}(k)$ for the P-ACT-LB (grey) and CMB-PAS datasets.  CMB-PAS is shown both with (cyan) and without (gold) DESI DR2 BAO data~\cite{DESI:2025zgx}. The official P-ACT-LB chains of~\cite{AtacamaCosmologyTelescope:2025nti} are shown in black.  \textit{Bottom:} The fractional uncertainty on $e^{-2\tau}\mathcal{P}(k)$ relative to the mean value for each bin. We see similar improvement of the CMB-PAS dataset compared to P-ACT-LB as in the seven-bin case shown in Figure~\ref{fig:primordialPk}. We also find that the addition of DESI DR2 BAO data does not significantly change the CMB-PAS constraints, as also see in Figure~\ref{fig:comparePk7bin} for the seven-bin case.}
    \label{fig:30bin}
    \label{fig:30bin}
    \end{minipage}%
    \hfill
    \begin{minipage}[t]{.49\textwidth}        \includegraphics[width=\textwidth]{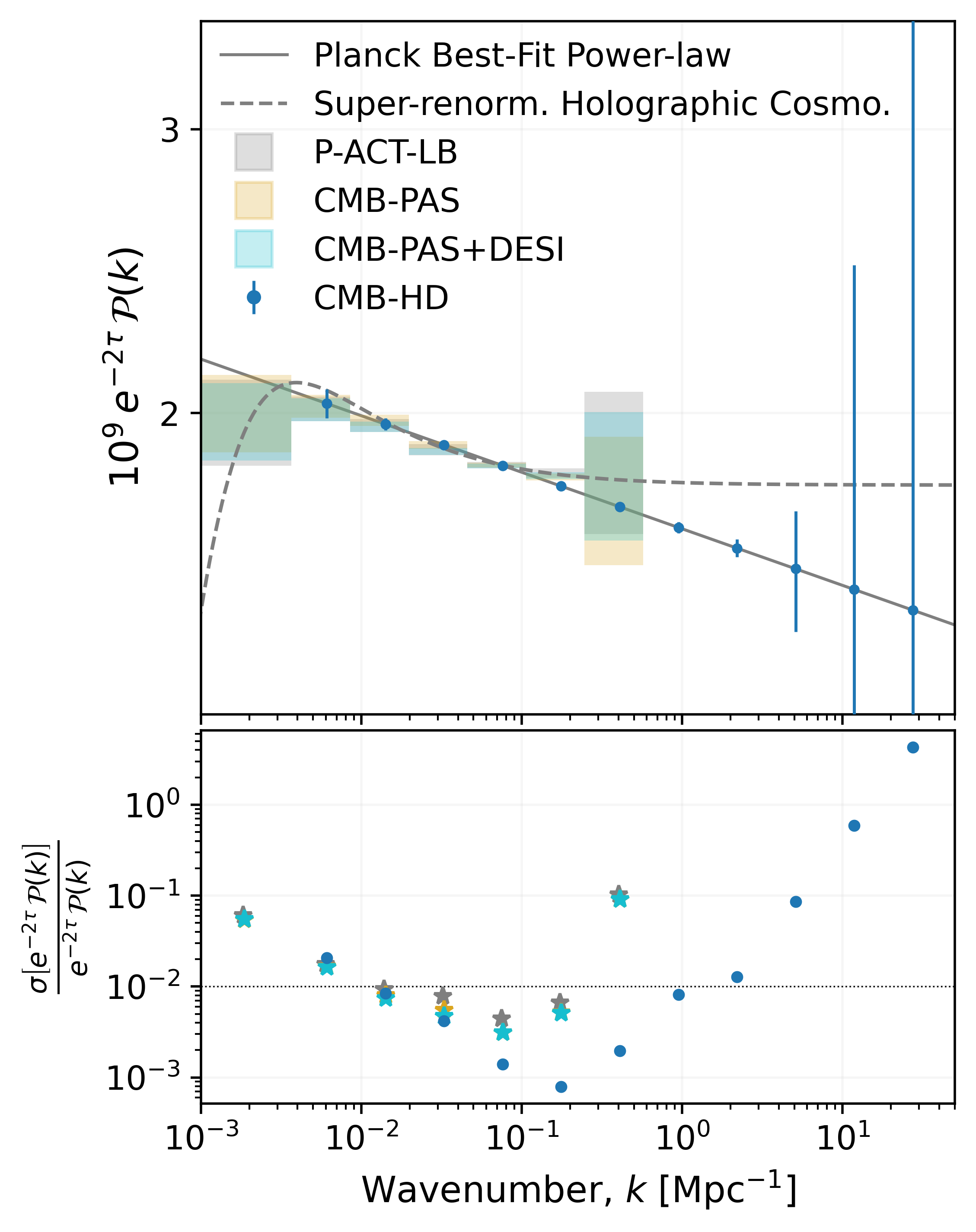}
    \caption{We show the effect of adding DESI DR2 BAO data to CMB-PAS on the binned primordial power spectrum, using the seven-bin scheme of Table~\ref{tab:P(k)}. \textit{Top:} Shown are marginalized means and $1\sigma$ error bars on $e^{-2\tau}\mathcal{P}(k)$ for each bin. \textit{Bottom:} The fractional uncertainty on $e^{-2\tau}\mathcal{P}(k)$ relative to the mean value.  We show P-ACT-LB (grey), CMB-PAS (gold), and CMB-PAS plus DESI DR2 (cyan).  We find little change in the CMB-PAS errors when adding DESI DR2 BAO. We also show forecasts for CMB-HD for comparison. }
\label{fig:comparePk7bin}
    \label{fig:comparePk7bin}
    \end{minipage}
\end{figure*}

\begin{table*}[t]
    \begin{center}
      \begin{tabular}{@{\hskip 1.5em} c@{\hskip 1.5em}  c@{\hskip 1.5em} c @{\hskip 1.5em} c @{\hskip 1.5em} c @{\hskip 1.5em} c @{\hskip 1.5em}}
      \toprule
      \hline
      \multicolumn{1}{c}{Bin Center} & \multicolumn{5}{c}{$10^9e^{-2\tau}\mathcal{P}(k)$}\\
\cmidrule(r){1-1} \cmidrule(l){2-6} 
      $k$  [Mpc$^{-1}$] & P-ACT-LB & CMB-PAS & CMB-PAS + DESI DR2 & SO-like & CMB-HD \\ 
      \midrule
  0.0018 & $1.992 \pm 0.113$ & $2.001 \pm 0.111$ & $1.978 \pm 0.110$ & --- & --- \\
      0.0060 & $2.013 \pm 0.036$ & $2.019 \pm 0.034$ & $2.009 \pm 0.033$ & $\pm 0.0428$ & $\pm 0.0419$ \\
      0.0141 & $1.964 \pm 0.018$ & $1.979 \pm 0.016$ & $1.961 \pm 0.015$ & $\pm 0.0156$ & $\pm 0.0167$ \\
      0.0327 & $1.900 \pm 0.012$ & $1.911 \pm 0.011$ & $1.893 \pm 0.009$ & $\pm 0.00839$ & $\pm 0.00799$ \\
      0.0759 & $1.856 \pm 0.009$ & $1.856 \pm 0.006$ & $1.854 \pm 0.006$ & $\pm 0.00274$ & $\pm 0.00262$ \\
      0.176 & $1.835 \pm 0.013$ & $1.825 \pm 0.010$ & $1.828 \pm 0.009$ & $\pm 0.00183$ & $\pm 0.00143$ \\
      0.408 & $1.769 \pm 0.206$ & $1.771 \pm 0.164$ & $1.834 \pm 0.167$ & $\pm 0.0196$ & $\pm 0.00347$ \\
      0.947 & --- & --- & --- & $\pm 0.154$ & $\pm 0.0139$ \\
      2.20 & --- & --- & --- & --- & $\pm 0.0210$ \\
      5.09 & --- & --- & --- & --- & $\pm 0.138$ \\
      11.8 & --- & --- & --- & --- & $\pm 0.920$ \\
      27.4 & --- & --- & --- & --- & $\pm 6.49$ \\
      \bottomrule
    \end{tabular}
    \caption{The marginalized means and $1\sigma$ errors on $e^{-2\tau}\mathcal{P}(k)$ for a binned primordial power spectrum from the P-ACT-LB (second column), CMB-PAS (third column), and CMB-PAS+DESI DR2 (fourth column) datasets. The first column displays the central wavenumber of the bin in $\mathrm{Mpc}^{-1}$. The CMB-PAS constraints are consistent with those of P-ACT-LB in all bins, with smaller error bars. Adding DESI DR2 to CMB-PAS minimally impacts parameter constraints.  We also show forecasts for SO-like and CMB-HD-like surveys, noting, in particular, that CMB-HD can go to $k$ values almost two orders of magnitude higher than current surveys.}
    \label{tab:pk-errors}
    \end{center}
\end{table*}

\clearpage 
\bibliographystyle{apsrev4-1}
\bibliography{main.bib}

\end{document}